\documentclass[final,5p,times,twocolumn]{elsarticle}

\usepackage{amssymb,amsmath}
\usepackage{graphicx}
\usepackage{booktabs,array,tabularx,float,titlesec}
\usepackage{microtype}
\usepackage{hyperref}
\usepackage{orcidlink}

\titleformat{\paragraph}[block]{\normalfont\normalsize\bfseries}{\theparagraph}{1em}{}
\titlespacing*{\paragraph}{0pt}{2ex plus .5ex}{.6ex}
\journal{Engineering Applications of Artificial Intelligence}

\begin{document}
\begin{frontmatter}

\title{Integrating Local Detail and Global Context: A Dual-Input Multi-Task Learning Framework for Bone Tumor Diagnosis}

\author[1]{S.M. Nasif Uddin\orcidlink{0009-0004-1522-9462}}
\author[2]{Rusab Sarmun\orcidlink{0009-0004-4887-0627}}
\author[3]{Muhammad E. H. Chowdhury\orcidlink{0000-0003-0744-8206}\corref{cor1}}
\author[4]{Adam Mushtak\orcidlink{0000-0001-6409-135X}}
\author[4]{Israa Al-Hashimi\orcidlink{0000-0001-7901-8841}}
\author[4]{Sohaib Bassam Zoghoul\orcidlink{0000-0001-9024-0670}}

\affiliation[1]{
    organization={Department of Electrical and Electronic Engineering, Ahsanullah University of Science and Technology},
    country={Bangladesh}
}

\affiliation[2]{
    organization={Department of Electrical and Electronic Engineering, University of Dhaka},
    country={Bangladesh}
}

\affiliation[3]{
    organization={Department of Electrical Engineering, Qatar University},
    addressline={Doha 2713},
    city={Doha},
    country={Qatar}
}

\affiliation[4]{
    organization={Department of Radiology, Hamad Medical Corporation},
    city={Doha},
    country={Qatar}
}

\cortext[cor1]{Corresponding author: Muhammad E.H. Chowdhury (mchowdhury@qu.edu.qa). Other author emails: S.M. Nasif Uddin (nasifuddinoff@gmail.com); Rusab Sarmun (rusabsarmun@gmail.com); Adam Mushtak (adamrads94@gmail.com); Israa Al-Hashimi (Ialhashimi@hamad.qa); Sohaib Bassam Zoghoul (sohaibzoghoul@gmail.com).}

\begin{abstract}
Primary bone tumors are rare but clinically aggressive neoplasms whose diagnosis from radiographs is challenged by heterogeneous morphology, subtle lesion margins, and overlapping bone structures. To address the limitations of existing single-view models, we present a dual-input, multi-task learning framework that, to our knowledge, is the first to apply bidirectional cross-modal attention between a lesion crop and the full radiograph for joint segmentation and subtype classification. Using the multi-institutional Bone Tumor X-ray Radiograph Dataset (BTXRD, n=3,746), we employ a YOLO-based detector to generate regions of interest, which are paired with full images as inputs to a dual-stream DenseNet121 architecture. Features are integrated via a novel cross-modal attention fusion strategy, refined by Hierarchical Multi-scale Feature Fusion, effectively balancing fine-grained lesion detail with global anatomical context. Evaluated on a held-out patient-level test split, the model demonstrates superior performance over single-input baselines, achieving an overall Dice Similarity Coefficient of 0.896 and a macro-averaged classification F1-score of 0.928. Notably, the system exhibits exceptional sensitivity for malignant osteosarcoma (AUC 0.999), validating the potential of dual-stream context modeling to support radiologists in accurate, early decision-making.
\end{abstract}
\begin{keyword}
bone tumour \sep radiograph \sep multi-task learning \sep cross-modal attention \sep semantic segmentation \sep computer-aided diagnosis
\end{keyword}
\end{frontmatter}

\section{Introduction}\label{introduction}

Primary malignant bone tumors, although relatively rare, are among the most aggressive musculoskeletal neoplasms. They rank as the third leading cause of cancer-related mortality in patients under the age of 20, with osteosarcoma and Ewing's sarcoma being particularly common subtypes \cite{ref1}. Accurate and timely diagnosis is crucial to improving survival rates and enabling limb-salvage surgeries. However, diagnosis based on radiographs is notoriously challenging. Bone tumors exhibit substantial heterogeneity in morphology, size, and radiodensity, while overlapping appearances with normal structures or benign abnormalities often obscure subtle features. This variability, combined with the limited field-of-view and 2D projection nature of radiographs, makes diagnosis difficult even for expert radiologists, leading to considerable inter-observer variability. Radiographs remain the most widely used imaging modality for bone tumor screening and initial evaluation because of their availability, low cost, and diagnostic value \cite{ref2}. Yet they are also among the most difficult to interpret, particularly in early stages of disease when lesions may occupy only a small portion of the image. More advanced imaging modalities such as MRI and CT provide richer information but are less accessible, more costly, and not always feasible in resource-limited settings. As such, radiographs represent an essential but imperfect tool in the diagnostic workflow, creating a compelling need for computer-aided diagnosis (CAD) systems that can enhance radiologist performance and provide consistent, reproducible results.

Recent advances in artificial intelligence (AI), particularly deep learning, have reshaped medical image analysis. Convolutional neural networks (CNNs) and transformer-based architectures have demonstrated state-of-the-art performance in tasks such as tumor segmentation, fracture detection, and lesion classification across multiple imaging modalities \cite{ref3,ref4,ref5}. In musculoskeletal oncology, deep learning--based radiomics has shown promise in extracting subtle quantitative features that may escape visual inspection, enabling more accurate tumor characterization \cite{ref6}. Despite these advances, the integration of CAD into clinical practice has been slow, largely due to limitations in available data and the complexity of modeling tumor heterogeneity. A major bottleneck in the development of CAD systems for bone tumors is the scarcity of large, well-annotated public datasets. The IIEST dataset, for example, contains only 69 bone tumor cases, which is insufficient for training high-capacity neural networks \cite{ref10}. Several institutional datasets have been developed, such as the collections by He et al. (2020) \cite{ref7}, von Schacky et al. (2021) \cite{ref8}, and Liu et al. \cite{ref9}. However, these are often restricted to specific anatomical regions, lack normal controls, or remain publicly inaccessible. Consequently, most prior works have relied on private datasets, hindering reproducibility and limiting the generalizability of models.

To address this gap, Yao et al. (2025) introduced the Bone Tumor X-ray Radiograph Dataset (BTXRD), a large-scale, multi-institutional resource comprising 3,746 radiographs annotated with bounding boxes, segmentation masks, and detailed clinical metadata. BTXRD includes normal, benign, and malignant cases across multiple anatomical regions, providing a foundation for classification, localization, and segmentation tasks. Importantly, each tumor image is accompanied by subtype labels and demographic information, reflecting clinically relevant features. Initial baseline experiments with YOLOv8 demonstrated the dataset's suitability for deep learning, paving the way for more advanced architectures \cite{ref10}. While BTXRD represents a significant step forward, it also highlights the challenges inherent to bone tumor CAD: tumors may be small relative to the image, morphologically diverse, and context-dependent, requiring models to balance fine-grained detail with broader anatomical understanding. Most prior CAD systems for bone tumors have approached the problem as a binary classification task, distinguishing benign from malignant lesions \cite{ref7,ref9}. While clinically useful, this oversimplification neglects the importance of subtype classification, which can guide treatment decisions and prognostic assessment. Furthermore, current architectural approaches face a fundamental dilemma. Single-input models trained on full radiographs typically rely on downsampling, which risks causing small lesions to vanish or blurring critical trabecular details. Conversely, models trained solely on cropped regions of interest (ROIs) preserve local texture but lose valuable anatomical context---such as the tumor\textquotesingle s location relative to the joint or the extent of periosteal reaction---which is vital for differential diagnosis. As a result, both approaches face limitations in generalization and robustness. Recent works have begun to explore multitask learning frameworks that combine segmentation and classification, showing that jointly optimizing for both objectives can improve feature learning\cite{ref9,ref11}. However, most of these models still rely on single-stream encoders that process either global or local views, rather than effectively integrating both.

In this study, we introduce a dual-input, multi-task learning framework that operates directly on radiographs to deliver both tumor segmentation and multi-class classification. The design explicitly preserves fine-grained lesion detail from cropped ROIs while retaining global skeletal context from full images, addressing the well-known trade-off between local precision and global interpretability in bone tumor CAD. Concretely, we adopt a CA-UNet--style decoder and employ two parallel encoders (for ROI and full image streams), then fuse their representations through cross-attention and hierarchical multi-scale feature fusion before branching to segmentation and classification heads. This mirrors the clinical workflow---localizing margins for surgery while inferring biological behavior---and aligns with recent evidence that multi-task objectives can regularize shared encoders and improve downstream performance when tasks are clinically complementary \cite{ref7,ref8,ref11,ref12}. Our fusion strategy follows the parallel-hybrid philosophy exemplified by HiFuse, which maintains distinct local/global streams and performs adaptive hierarchical fusion rather than forcing early abstraction \cite{ref13}. To further sharpen discriminative features, we integrate lightweight attention in the encoder/decoder and fusion blocks via CBAM to emphasize task-relevant channels and spatial regions while suppressing noise---an approach shown to enhance medical image segmentation with minimal overhead \cite{ref14}. Because radiographic tumors often occupy a small fraction of pixels, we train with an imbalance-aware objective, adopting Unified Focal Loss to combine region-overlap sensitivity (Dice) with hard-example focusing (Focal), which has demonstrated robust gains across class-imbalanced medical segmentation benchmarks \cite{ref15}. We position our framework relative to two influential directions: (i) hybrid CNN--Transformer architectures that inject global self-attention into U-shaped designs to capture long-range dependencies \cite{ref16,ref17} and (ii) modernized CNNs that achieve large receptive fields through Transformer-inspired convolutions and scaling, retaining CNN inductive biases and data-efficiency \cite{ref18}. While both families are promising, we deliberately do \textbf{not} replace our CNN encoder with MedNeXt nor migrate to Swin blocks in this work; instead, we test the hypothesis that explicit dual-stream (local+global) processing with learned cross-attention fusion offers a strong and interpretable solution for heterogeneous bone tumors on radiographs, complementing these alternatives.

\begin{figure*}[!t]
\centering
\includegraphics[width=0.94\textwidth,keepaspectratio]{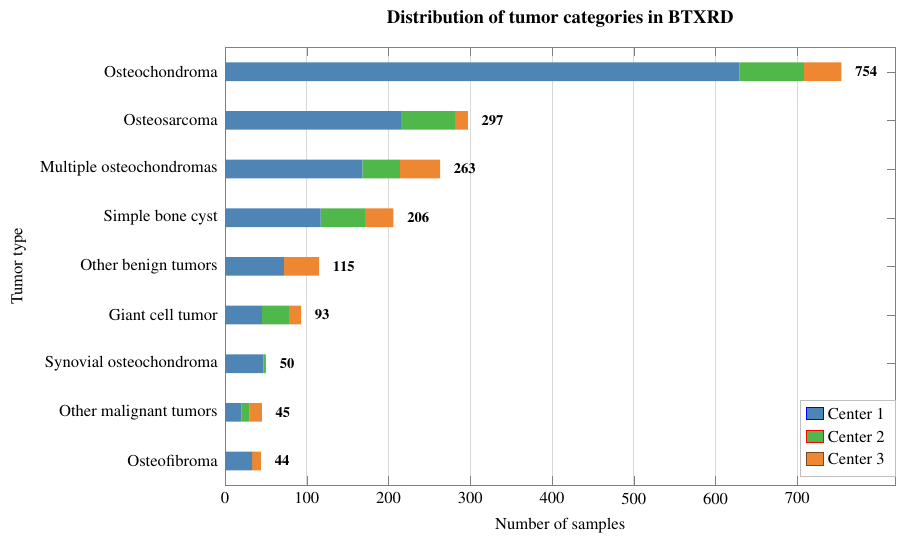}
\caption{Class distribution of the BTXRD dataset}
\label{fig:class-distribution}
\end{figure*}

The contributions of this study are the following:

\begin{itemize}
\item
  Established a dual-input, multi-task learning paradigm for bone tumor diagnosis that simultaneously performs segmentation and multi-class classification. Unlike single-stream baselines, this framework explicitly integrates fine-grained lesion details from cropped ROIs with global skeletal context from full radiographs, effectively addressing the trade-off between local precision and global interpretability.
\item
  Proposed a cross-modal attention fusion strategy that dynamically aligns local and global feature streams. By incorporating hierarchical multi-scale fusion and self-attention refinement, the architecture preserves spatial relationships critical for delineating small or morphologically complex tumors while maintaining semantic consistency for classification.
\item
  Incorporated a self-attention block on fused features to model intra-tumor dependencies and refine the learned representations.
\item
  Demonstrated superior diagnostic performance on the large-scale, multi-institutional BTXRD dataset. The model extends beyond binary categorization to achieve high sensitivity in clinically relevant subtypes, such as osteosarcoma and osteochondroma, validating the efficacy of dual-stream context modeling without reliance on multimodal data.
\end{itemize}

\section{Methodology}\label{methodology}

This section follows the pipeline in the order in which it is applied. Section 2.1 introduces the BTXRD dataset and the three-class taxonomy adopted here. Section 2.2 covers preprocessing: consolidation of benign subtypes, automated tumour localisation with a YOLOv11 detector and extraction of standardised regions of interest, the patient-level data split, and the augmentation pipeline. Section 2.3 presents OsteoHiFuse-Net itself --- the two DenseNet121 encoding streams, the bidirectional cross-modal attention that fuses them at the bottleneck, the hierarchical multi-scale fusion module feeding the classifier, the attention-gated decoder feeding the segmentation head, and the deep-supervision scheme. Section 2.4 defines the compound multi-task objective, and Section 2.5 lists the segmentation and classification metrics used throughout the results.

\subsection{Dataset Description}\label{dataset-description}

\subsubsection{BTXRD Overview}\label{btxrd-overview}

This study utilized the Bone Tumor X-ray Radiograph Dataset (BTXRD), a large-scale, multi-institutional dataset recently introduced by Yao et al. (2025) \cite{ref10}. BTXRD comprises 3,746 radiographs collected from five medical institutions, representing one of the most comprehensive publicly available resources for bone tumor analysis. The dataset includes both normal bone radiographs (n=1,879) and tumor cases (n=1,867), with tumor samples further categorized into benign (n=1,525) and malignant (n=342) subtypes. Each radiograph is accompanied by expert annotations including bounding boxes for tumor localization, pixel-level segmentation masks, and detailed clinical metadata such as patient age, gender, anatomical location, and histopathological diagnosis.

The dataset spans multiple anatomical sites, with the most common locations being the tibia (630 lesions), femur (592 lesions), fibula (259 lesions), and humerus (235 lesions), reflecting the clinical distribution of primary bone tumors. All images were acquired using standard radiographic protocols, capturing the heterogeneity of imaging equipment and acquisition parameters across different institutions. This multi-institutional nature enhances the generalizability of models trained on BTXRD, as they must learn robust features invariant to site-specific imaging characteristics.

The tumor class distribution is shown in Figure~\ref{fig:class-distribution}.

\subsubsection{Class Distribution and Clinical Relevance}\label{class-distribution-and-clinical-relevance}

To address clinically relevant diagnostic challenges while maintaining statistical robustness, we focused on three specific tumor categories:

\begin{itemize}
\item
  \textbf{Osteosarcoma:} The most common primary malignant bone tumor, requiring timely intervention for limb salvage.
\item
  \textbf{Osteochondroma:} The most prevalent benign tumor, necessitating accurate identification to prevent unnecessary invasive procedures.
\item
  \textbf{Other Benign:} A consolidated category comprising Giant Cell Tumors and Simple Bone Cysts.
\end{itemize}

\subsection{Data Preprocessing Pipeline}\label{data-preprocessing-pipeline}

\subsubsection{Class Consolidation Strategy}\label{class-consolidation-strategy}

The original BTXRD dataset contains multiple fine-grained benign tumor subtypes with highly imbalanced distributions. To ensure robust model training and clinical applicability, we consolidated the giant cell tumour and simple bone cyst subtypes into a single "other benign" category and excluded the remaining, severely underrepresented subtypes. This grouping introduces some intra-class heterogeneity, but it mirrors the clinical management pathway, in which establishing that a lesion is non-malignant is the primary decision node before specific subtyping. This three-class taxonomy---osteosarcoma, osteochondroma, and other benign---was designed to reflect diagnostic decisions that directly impact clinical management: distinguishing malignant tumors requiring aggressive treatment from benign lesions that may be managed conservatively, while also identifying the specific benign subtype most commonly encountered in clinical practice.

This consolidation reduced the risk of model overfitting to underrepresented classes and improved training stability by ensuring each class had sufficient examples for the model to learn generalizable features. The final class distribution after consolidation but before augmentation reflected the natural clinical prevalence of these tumor types, with benign tumors substantially outnumbering malignant cases, consistent with epidemiological data.

\subsubsection{Automated Tumor Localization and ROI Generation}\label{automated-tumor-localization-and-roi-generation}

To enable dual-input processing, tumor regions must first be localized within full radiographs and cropped into standardized regions of interest (ROIs). Manual annotation of bounding boxes is time-consuming and subject to inter-observer variability; therefore, we employed an automated detection pipeline using YOLOv11 \cite{ref19}, a state-of-the-art real-time object detection architecture.

Crucially, to prevent data leakage, the object detector was trained strictly on the training split of the dataset. No images from the validation or test sets were seen by the YOLO model during its training phase. The detector locates the tumor and generates a bounding box, which is then used to crop the ROI for the downstream segmentation and classification network. This two-stage pipeline mirrors a realistic clinical deployment where manual cropping is not feasible.

YOLOv11 is a single-stage object detector that performs localization and classification in a unified forward pass, making it computationally efficient for clinical deployment. The architecture employs a CSPDarknet backbone for feature extraction, a Path Aggregation Network (PANet) for multi-scale feature fusion, and anchor-free detection heads that predict bounding boxes and class probabilities directly. This design is particularly well-suited for medical imaging, where tumors exhibit substantial scale variation and require precise localization.

The YOLOv11 model was trained on the full BTXRD training set using the bounding box annotations provided with the dataset. Training was conducted using the Ultralytics framework with the configuration in Table~\ref{tab:yolo-config}.

\begin{table}[!t]
\centering
\caption{YOLOv11 training configuration.}
\label{tab:yolo-config}
\small
\begin{tabular}{@{}ll@{}}
\toprule
Parameter & Value \\
\midrule
Model Variant & YOLOv11x \\
Input Resolution & $640\times640$ \\
Batch Size & 8 \\
Training Epochs & 118 \\
Optimizer & SGD \\
Initial Learning Rate & $1\times10^{-3}$ \\
Mosaic Augmentation & Enabled \\
HSV Augmentation & $H=0.015$, $S=0.7$, $V=0.4$ \\
Horizontal Flip & $p=0.5$ \\
Rotation & $\pm10^\circ$ \\
Scale Augmentation & $0.5$--$1.5\times$ \\
\bottomrule
\end{tabular}
\end{table}

The model was initialized with COCO-pretrained weights and fine-tuned on BTXRD for bone tumor-specific feature learning. Training employed automatic mixed precision (AMP) for computational efficiency and was terminated using early stopping when validation mean average precision (mAP) plateaued for 20 consecutive epochs.

The trained YOLOv11 detector was evaluated on the validation set to assess localization accuracy before using it for ROI extraction. We measured precision, recall, mAP@0.5, and mAP@0.5:0.95.

Detected bounding boxes from YOLOv11 were used to extract standardized 256\ensuremath{\times}256 pixel ROI crops according to the following algorithm:

Detected bounding boxes were converted into standardised ROI crops as follows. Each box was first expanded by 10\% in all directions, subject to the image boundary, so that peripheral features such as periosteal reaction and soft-tissue involvement were retained. The expanded box was then squared by taking the larger of its width and height and re-centring on the original detection centroid, giving a uniform aspect ratio without discarding surrounding anatomy. The square region was finally resampled to 256 \ensuremath{\times} 256 pixels by bilinear interpolation, a resolution that preserves sufficient texture and boundary detail while keeping the dual-input forward pass tractable.

Two edge cases were handled explicitly. Lesions smaller than 256 pixels in their longest dimension were centred within the crop and padded in reflection mode to reach the target size, avoiding the artificial edges introduced by zero padding; lesions larger than 256 pixels were downscaled so that the whole lesion remained visible. Where the detector returned more than one box for a radiograph, each detection produced an independent crop-and-full-image pair, allowing multifocal cases to be processed without modification. As a quality-control step, crops with a detection confidence below 0.5 were reviewed manually to remove false positives, and crops failing an anatomical plausibility check (for example, boxes covering less than 5\% tumour pixels) were excluded.

Each cropped ROI was paired with its corresponding full radiograph to form a dual-input sample. This pairing gives the network simultaneous access to high-resolution local detail from the 256 \ensuremath{\times} 256 crop --- fine-grained tumour morphology, matrix mineralisation patterns and margin characteristics --- and to the global anatomical context of the full radiograph, which carries periosteal reaction, cortical involvement, soft-tissue extension and the spatial relationship of the lesion to adjacent structures.

The ROI extraction pipeline was applied identically to training, validation, and test sets, with all subsequent experiments using these pre-generated crops paired with their source full images.

\subsubsection{Train--Validation Split}\label{trainvalidation-split}

The dataset was divided into training (80\%), validation (10\%), and testing (10\%) sets using a stratified random split. We enforced strict patient-level separation, ensuring that all radiographs from a single patient were assigned to the same subset. This prevents the model from memorizing patient-specific anatomical features (e.g., bone structure or implants) that could lead to inflated performance metrics. The augmentation pipeline was applied after this split, exclusively to the training data.

\subsubsection{Preprocessing and Augmentation}\label{preprocessing-and-augmentation}

All cropped ROIs were resized to a uniform dimension of 256\ensuremath{\times}256 pixels to ensure consistent input dimensions for the neural network. Full radiographs were resized to 256x256 pixels while preserving the original aspect ratio through padding. To enhance model generalization and mitigate class imbalance, we applied a diverse set of medically relevant augmentations only to the training subset. The validation and test sets remained unaltered to ensure an unbiased assessment of model performance. Each transformation was carefully selected to maintain anatomical plausibility and preserve diagnostic features while introducing realistic variability in acquisition conditions.

Geometric transformations simulated variation in patient positioning and device calibration. Images were rotated by up to \ensuremath{\pm}15\ensuremath{^\circ} (p = 0.6), a range chosen to remain anatomically plausible, and flipped horizontally with p = 0.5, which is appropriate for bone radiographs because left--right mirroring does not alter tumour morphology; vertical flips were applied more sparingly (p = 0.2), since vertical inversion is uncommon in clinical imaging. Elastic deformation (\ensuremath{\alpha} = 50, \ensuremath{\sigma} = 5, \ensuremath{\alpha}\_affine = 5, p = 0.3) modelled soft-tissue variability and minor non-rigid distortion, while grid distortion (5 steps, limit = 0.1, p = 0.2) reproduced geometric calibration differences between radiographic devices.

Intensity transformations accounted for differences in exposure and detector response. Brightness and contrast were jittered by up to \ensuremath{\pm}20\% (p = 0.6) to reflect exposure variation across imaging systems, and gamma correction with \ensuremath{\gamma} \ensuremath{\in} {[}0.8, 1.2{]} (p = 0.3) modelled the non-linear intensity mappings common in radiography.

Finally, noise and artefact simulation improved robustness to acquisition quality: Gaussian noise with variance in {[}10, 50{]} (p = 0.3) approximated electronic noise from digital detectors, and Gaussian blur with a kernel of at most 3 (p = 0.2) reproduced mild motion blur or focal shift.

All augmentations were implemented using the Albumentations library \cite{ref30}, ensuring consistent and reproducible transformations. Corresponding segmentation masks were identically transformed to maintain pixel-wise alignment.

\begin{figure}[H]
\centering
\includegraphics[width=\columnwidth,keepaspectratio]{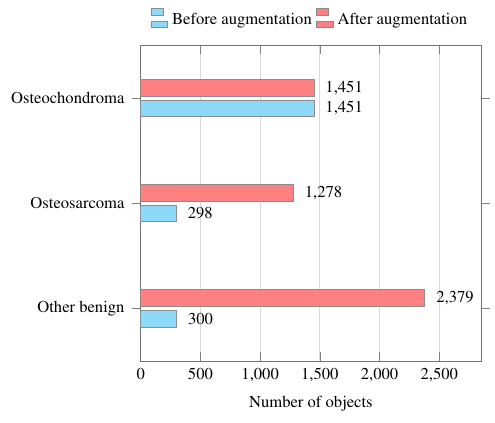}
\caption{Training-set class distribution before and after augmentation}
\label{fig:augmentation-distribution}
\end{figure}

After augmentation, class-specific sampling was used to reduce imbalance within the training data. The resulting distribution consisted of \textbf{5108} samples, as shown in \emph{Figure~\ref{fig:augmentation-distribution}}. The class counts are listed below.

\begin{itemize}
\item
  \textbf{O}steochondroma: 1451 (28.4\%)
\item
  \textbf{O}steosarcoma: 1278 (25.0\%)
\item
  \textbf{O}ther benign: 2379 (46.6\%)
\end{itemize}

This adjustment produced an approximately 1 : 1 : 2 ratio (osteochondroma : osteosarcoma : other benign)---substantially improving balance while retaining the natural prevalence hierarchy (benign \textgreater{} malignant). Figure~\ref{fig:augmentation-distribution} illustrates the class composition before and after augmentation.

\subsection{Proposed Dual-Input Multi-Task Architecture}\label{proposed-dual-input-multi-task-architecture}

\begin{figure*}[!t]
\centering
\includegraphics[width=0.88\textwidth,height=0.66\textheight,keepaspectratio]{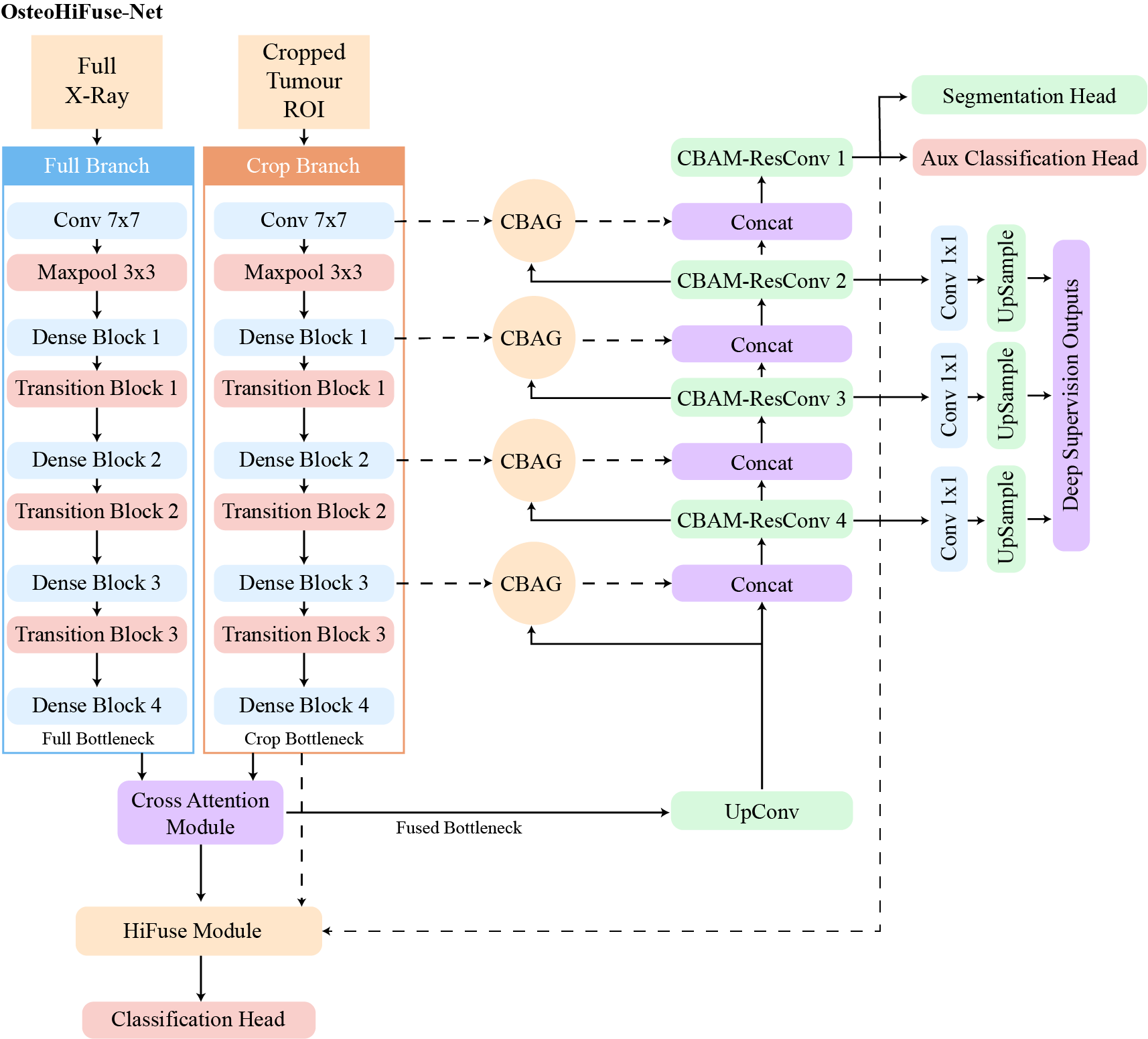}
\caption{Proposed dual-input multi-task OsteoHiFuse-Net architecture}
\label{fig:architecture}
\end{figure*}

Modern bone tumor diagnosis from radiographs requires the integration of two complementary types of information: fine-grained lesion morphology visible in cropped tumor regions, and broader anatomical context from full radiographs that reveals the tumor\textquotesingle s relationship to surrounding skeletal structures. Single-input approaches sacrifice one of these perspectives---models trained on full radiographs may overlook subtle lesion features, while crop-only models lose critical contextual cues such as periosteal reaction patterns, bone remodeling, and spatial relationships that inform differential diagnosis. To address this fundamental trade-off, we propose OsteoHiFuse-Net, a dual-input attention-guided architecture that explicitly processes both perspectives through separate encoding pathways and fuses them through learnable cross-modal attention mechanisms.

\emph{Figure~\ref{fig:architecture}} illustrates the complete architecture. Its five major components are (1) dual-stream CNN encoders with progressive attention enhancement, (2) cross-modal attention fusion at the bottleneck, (3) hierarchical multi-scale feature integration for classification, (4) an attention-gated U-Net decoder for segmentation, and (5) dual-task output heads with deep supervision. The design philosophy follows recent evidence that maintaining separate local and global feature streams until late fusion preserves scale-specific information while enabling adaptive integration through attention, in contrast to early fusion approaches that prematurely collapse multi-scale representations.

\subsubsection{Overall Framework Design}\label{overall-framework-design}

OsteoHiFuse-Net implements a parallel-encoder, shared-decoder architecture optimized for multi-task learning. Two independent encoding pathways process the cropped tumor ROI (256\ensuremath{\times}256 pixels) and full radiograph separately, extracting hierarchical feature representations from low-level edges and textures to high-level semantic patterns. At the bottleneck level, features from both streams are aligned and fused via a cross-modal attention module that computes bidirectional attention weights, allowing crop features to query relevant global context and vice versa. This attention-weighted fusion generates an enriched bottleneck representation that combines fine-grained local detail with broader spatial context.

The fused bottleneck features are then passed to a U-Net-style decoder for segmentation, where skip connections from the crop encoder are modulated by enhanced attention gates to suppress irrelevant features and emphasize diagnostically important regions. Simultaneously, a hierarchical feature fusion module aggregates multi-scale representations from the crop bottleneck, fused bottleneck, and final decoder features to generate a compact embedding for classification. This multi-scale aggregation ensures that the classification head receives information spanning from low-level texture (decoder features) to high-level semantics (bottleneck features), improving discriminative power for morphologically heterogeneous tumors.

The architecture outputs three predictions: (1) a pixel-wise tumor segmentation mask, (2) a three-class tumor type classification, and (3) an auxiliary classification from the crop bottleneck. The auxiliary classifier provides an additional gradient pathway directly from the classification loss to the crop encoder, which has been shown to improve feature learning in multi-task frameworks \cite{ref8,ref11}. Deep supervision is applied at three intermediate decoder stages, providing auxiliary segmentation signals that guide feature learning at multiple spatial scales.

The proposed architecture is not over-engineered but rather a purposeful response to the unique diagnostic challenge of bone tumors, which require the simultaneous interpretation of fine-grained lesion morphology and broader skeletal context. Single-stream models typically face a fundamental trade-off where they either lose subtle lesion features through down\textbf{-}sampling full radiographs or lose vital anatomical context---such as periosteal reactions and spatial relationships---through cropping. Among the various components, the dual-stream configuration (DenseNet \ensuremath{\times} 2) coupled with cross-modal attention fusion is the most critical element, as it provides the foundational structural mechanism for integrating these two essential perspectives. While lightweight modules like CBAM and deep supervision provide necessary refinement and training stability with minimal overhead, they act as supporting features for the core innovation of the dual-input framework, which is what ultimately enables the system to outperform single-view baselines and achieve exceptional sensitivity for malignant subtypes such as osteosarcoma.

\subsubsection{Encoder Architecture}\label{encoder-architecture}

\paragraph{DenseNet121 Backbone Selection}\label{densenet121-backbone-selection}

Both the crop and full-image encoding pathways utilize DenseNet121 \cite{ref27} as the base architecture, chosen for its dense connectivity pattern that facilitates gradient flow and feature reuse---properties particularly valuable for medical imaging where training data is limited and hierarchical feature propagation is critical. DenseNet121\textquotesingle s architecture connects each layer to all subsequent layers through concatenation, creating implicit deep supervision that mitigates vanishing gradients and enables efficient parameter utilization.

The encoders were initialized with ImageNet-pretrained weights to leverage transfer learning from natural images, a strategy that has proven effective even for modalities as distinct as X-ray radiographs \cite{ref7,ref9}. Transfer learning provides a strong initialization that captures general visual features (edges, textures, shapes) and accelerates convergence, particularly important given the limited size of medical datasets relative to natural image collections.

\begin{figure*}[!t]
\centering
\includegraphics[width=0.88\textwidth,height=0.66\textheight,keepaspectratio]{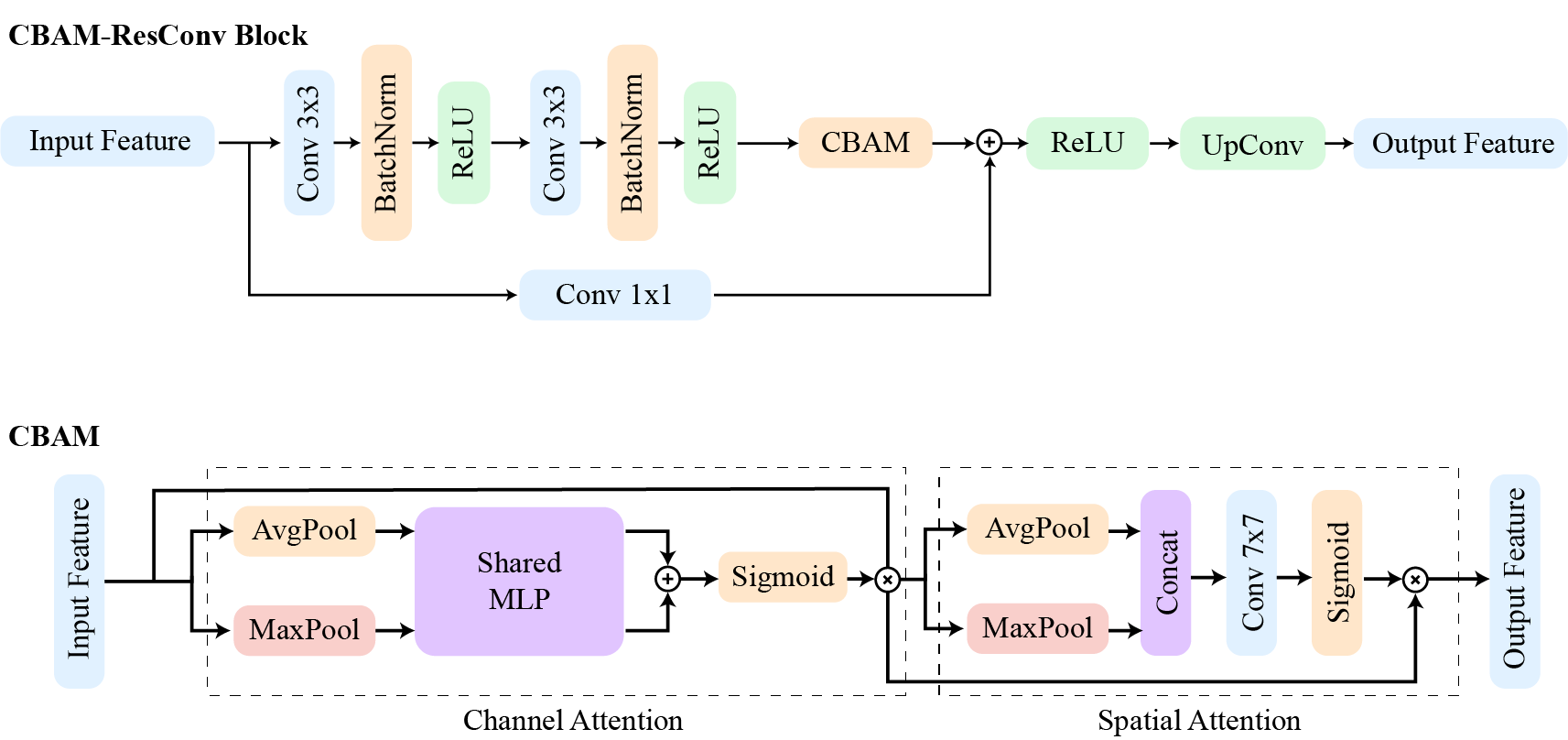}
\caption{Convolutional Block Attention Module (CBAM).}
\label{fig:cbam}
\end{figure*}

\paragraph{Dual-Stream Configuration}\label{dual-stream-configuration}

We employ two independent DenseNet121 encoders---one for the cropped ROI and one for the full radiograph (see \emph{Figure~\ref{fig:architecture}})---rather than a single shared encoder. This design choice preserves scale-specific feature learning: the crop encoder learns to recognize subtle intra-tumoral patterns (trabecular destruction, matrix mineralization, margin characteristics) at high spatial resolution, while the full encoder captures coarser anatomical relationships (periosteal reaction, bone expansion, soft tissue involvement) at lower effective resolution. Sharing a single encoder would force the model to learn a compromise representation that may be suboptimal for both scales.

Using the timm library, we extract multi-scale feature maps from five hierarchical stages of each encoder. For DenseNet121, these stages produce feature maps with channel dimensions (after the transition layers) and progressively decreasing spatial. However, our implementation reports enhanced channel counts due to feature extraction at different intermediate points optimized for medical imaging applications.

\paragraph{Progressive Feature Enhancement}\label{progressive-feature-enhancement}

To further improve the discriminative power of encoder features, each stage\textquotesingle s output is passed through a progressive feature enhancement module consisting of:

\begin{itemize}
\item
  Depthwise separable convolution \cite{ref20}: A 3\ensuremath{\times}3 depthwise convolution (operating independently on each channel) followed by a 1\ensuremath{\times}1 pointwise convolution. This factorization reduces computational cost while maintaining expressive power and has proven effective for capturing spatial patterns in medical images.
\item
  Batch normalization and ReLU activation: Stabilizes training dynamics and introduces non-linearity for feature refinement.
\end{itemize}

The enhanced features from each stage are then passed to CBAM attention modules (described below) before being used for skip connections or further downsampling.

\paragraph{CBAM Attention Integration}\label{cbam-attention-integration}

Convolutional Block Attention Module (CBAM) \cite{ref14} is integrated at every encoder stage to recalibrate feature responses both channel-wise and spatially (see \emph{Figure~\ref{fig:cbam}}). CBAM operates in two sequential stages.

Channel Attention Module: Computes a channel descriptor by aggregating spatial information through both average pooling and max pooling, producing two context vectors that are independently processed through a shared two-layer MLP (with reduction ratio 16) and summed. The resulting channel attention weights are applied via element-wise multiplication to emphasize informative feature channels while suppressing less relevant ones. Mathematically:

\begin{equation}
M_c(\mathbf F)=\sigma\!\left(W_1\!\left(W_0(\operatorname{AvgPool}(\mathbf F))\right)+W_1\!\left(W_0(\operatorname{MaxPool}(\mathbf F))\right)\right)
\end{equation}

where \(\mathbf{W}_{0}\in\mathbb{R}^{C/r \times C}\) and \(\mathbf{W}_{1}\in\mathbb{R}^{C \times C/r}\) are MLP weights with reduction ratio \(r = 16\), and \(\sigma \) is the sigmoid function.

Spatial Attention Module: After channel recalibration, spatial attention is computed by aggregating channel information through average and max pooling along the channel dimension, concatenating the two resulting 2D maps, and processing them with a 7\ensuremath{\times}7 convolution followed by sigmoid activation. The spatial attention map identifies "where" to focus, assigning higher weights to tumor regions and suppressing background. Mathematically:

\begin{equation}
M_s(\mathbf F)=\sigma\!\left(\operatorname{Conv}_{7\times7}\!\left([\operatorname{AvgPool}_c(\mathbf F);\operatorname{MaxPool}_c(\mathbf F)]\right)\right)
\end{equation}

\emph{\hfill\break
}

\begin{equation}
\mathbf F'=M_s(M_c(\mathbf F)\odot\mathbf F)\odot(M_c(\mathbf F)\odot\mathbf F)
\end{equation}

The final refined feature map is the input modulated by both attention maps in sequence.

CBAM was selected over alternatives like Squeeze-and-Excitation (SE) modules because it explicitly models spatial attention in addition to channel attention, which is crucial for medical segmentation where precise spatial localization of tumor boundaries is required. CBAM has demonstrated consistent improvements in medical imaging tasks with minimal computational overhead (\textless1\% parameter increase) \cite{ref14}.

\subsubsection{Cross-Modal Attention Fusion Module}\label{cross-modal-attention-fusion-module}

\begin{figure*}[!t]
\centering
\includegraphics[width=0.88\textwidth,height=0.66\textheight,keepaspectratio]{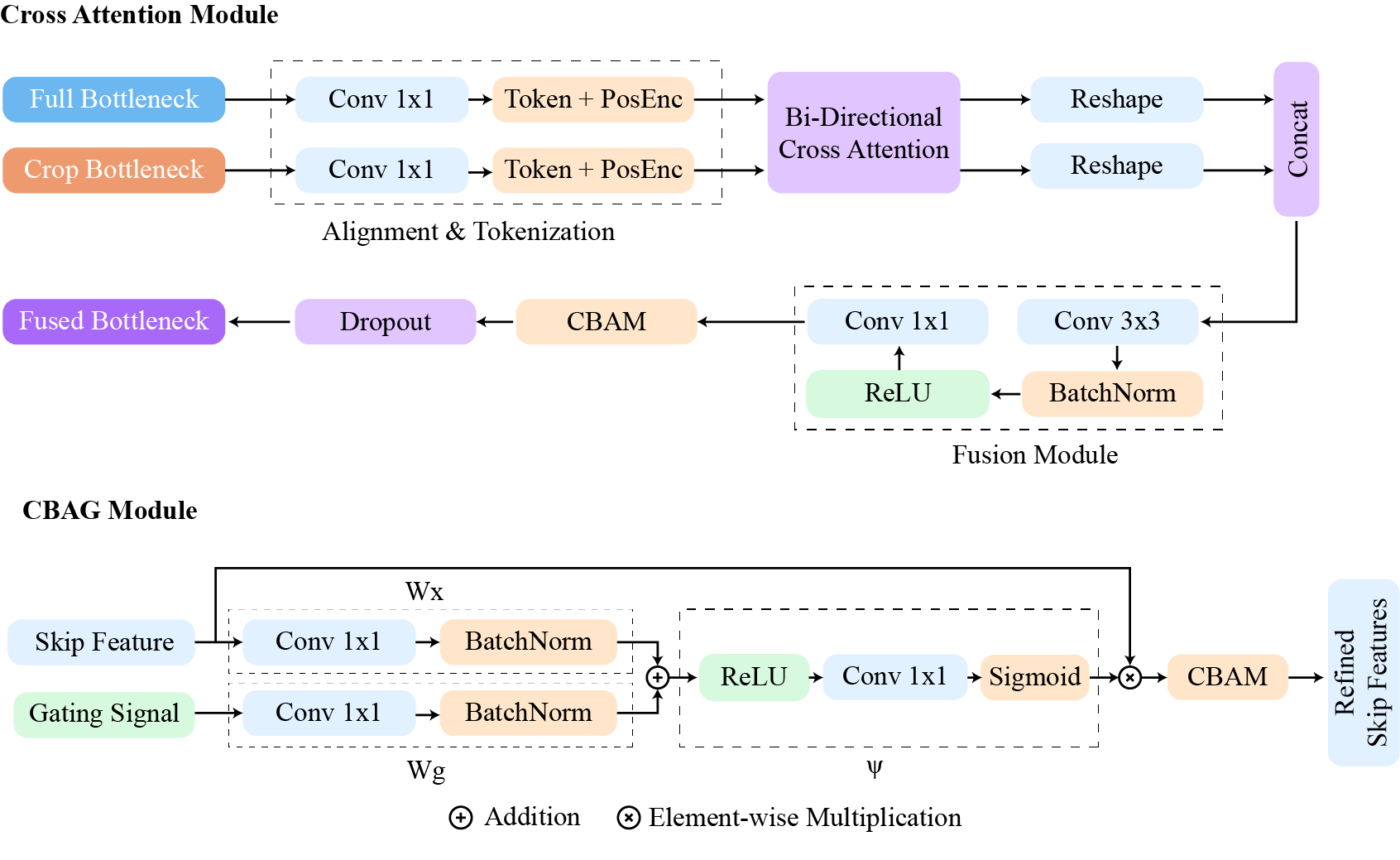}
\caption{Cross-modal attention fusion module.}
\label{fig:cross-modal-attention}
\end{figure*}

At the deepest encoder level (bottleneck), feature maps from the crop and full-image streams must be integrated to produce a unified representation that leverages both local detail and global context. Simple concatenation or element-wise addition would fail to model the semantic relationships between these two views. Instead, we employ a cross-modal attention mechanism inspired by multi-head self-attention \cite{ref21} but adapted for multi-scale feature fusion. Figure~\ref{fig:cross-modal-attention} depicts this module.

\paragraph{Query-Key-Value Architecture}\label{query-key-value-architecture}

The cross-modal attention module first projects both crop and full bottleneck features to a common embedding space of dimension 512 using 1\ensuremath{\times}1 convolutions:

\begin{equation}
\mathbf F_{\mathrm{crop}}^{\mathrm{proj}}=\operatorname{Conv}_{1\times1}(\mathbf F_{\mathrm{crop}}^{\mathrm{btl}})\in\mathbb R^{B\times512\times H\times W}
\end{equation}

\begin{equation}
\mathbf F_{\mathrm{full}}^{\mathrm{proj}}=\operatorname{Conv}_{1\times1}(\mathbf F_{\mathrm{full}}^{\mathrm{btl}})\in\mathbb R^{B\times512\times H\times W}
\end{equation}

where \(\mathbf{F}_{\text{crop}}^{\text{btl}}\) and \(\mathbf{F}_{\text{full}}^{\text{btl}}\) are the raw bottleneck features (320 channels for DenseNet121), and \(H \times W\) is the spatial dimension at the bottleneck.

These projected features are then reshaped from spatial format \((B,C,H,W)\) to sequence format \((B,H \cdot W,C)\) to enable self-attention computation. We perform bidirectional cross-attention:

Crop-to-Full Attention: Crop features serve as queries, full features as keys and values. This allows each spatial location in the crop to attend to relevant regions in the full image, capturing how local tumor details relate to global anatomical context.

\begin{equation}
\begin{aligned}\operatorname{Attended}_{\mathrm{crop}}={}&\operatorname{MultiHeadAttn}\bigl(Q=\mathbf F_{\mathrm{crop}}^{\mathrm{proj}},\\[-2pt]&K=\mathbf F_{\mathrm{full}}^{\mathrm{proj}},\;V=\mathbf F_{\mathrm{full}}^{\mathrm{proj}}\bigr)\end{aligned}
\end{equation}

\emph{\hfill\break
}

Full-to-Crop Attention: Full features query the crop, allowing the global view to extract detailed information from the tumor region.

\begin{equation}
\begin{aligned}\operatorname{Attended}_{\mathrm{full}}={}&\operatorname{MultiHeadAttn}\bigl(Q=\mathbf F_{\mathrm{full}}^{\mathrm{proj}},\\[-2pt]&K=\mathbf F_{\mathrm{crop}}^{\mathrm{proj}},\;V=\mathbf F_{\mathrm{crop}}^{\mathrm{proj}}\bigr)\end{aligned}
\end{equation}

\emph{\hfill\break
}

The multi-head attention mechanism (with 8 heads) computes scaled dot-product attention:

\begin{equation}
\operatorname{Attention}(Q,K,V)=\operatorname{softmax}\!\left(\frac{QK^{\mathsf T}}{\sqrt{d_k}}\right)V
\end{equation}

\emph{\hfill\break
}

where \(d_{k}= 512/8 = 64\) is the dimension per head. Multiple heads allow the model to jointly attend to different aspects of the feature space (e.g., one head might focus on bone density, another on margin characteristics).

\paragraph{Adaptive Feature Fusion}\label{adaptive-feature-fusion}

After bidirectional attention, the two attended feature maps are concatenated along the channel dimension (yielding 1024 channels), then fused through a convolutional refinement block:

\begin{equation}
\begin{aligned}\mathbf F_{\mathrm{fused}}={}&\operatorname{CBAM}\bigl(\operatorname{Conv}_{1\times1}(\operatorname{ReLU}(\\[-2pt]&\operatorname{BN}(\operatorname{Conv}_{3\times3}([\mathbf F_{\mathrm{crop}}^{\mathrm{att}};\mathbf F_{\mathrm{full}}^{\mathrm{att}}]))))\bigr)\end{aligned}
\end{equation}

This fusion block consists of:

\begin{itemize}
\item
  3\ensuremath{\times}3 convolution (512 channels) for spatial integration
\item
  Batch normalization and ReLU
\item
  1\ensuremath{\times}1 convolution (512 channels) for channel-wise refinement
\item
  CBAM attention for final feature recalibration
\item
  Dropout (0.1) for regularization
\end{itemize}

The resulting fused bottleneck representation \(\mathbf{F}_{\text{fused}}\in\mathbb{R}^{B \times 512 \times H \times W}\) serves as the input to the decoder and contains information from both encoding streams, dynamically weighted by their relevance.

The attention weights computed during crop-to-full attention are also returned for optional visualization, enabling interpretability analysis of which full-image regions the model attends to when processing each crop location.

\subsubsection{Hierarchical Feature Fusion (HiFuse) Module}\label{hierarchical-feature-fusion-hifuse-module}

\begin{figure*}[!t]
\centering
\includegraphics[width=0.88\textwidth,height=0.66\textheight,keepaspectratio]{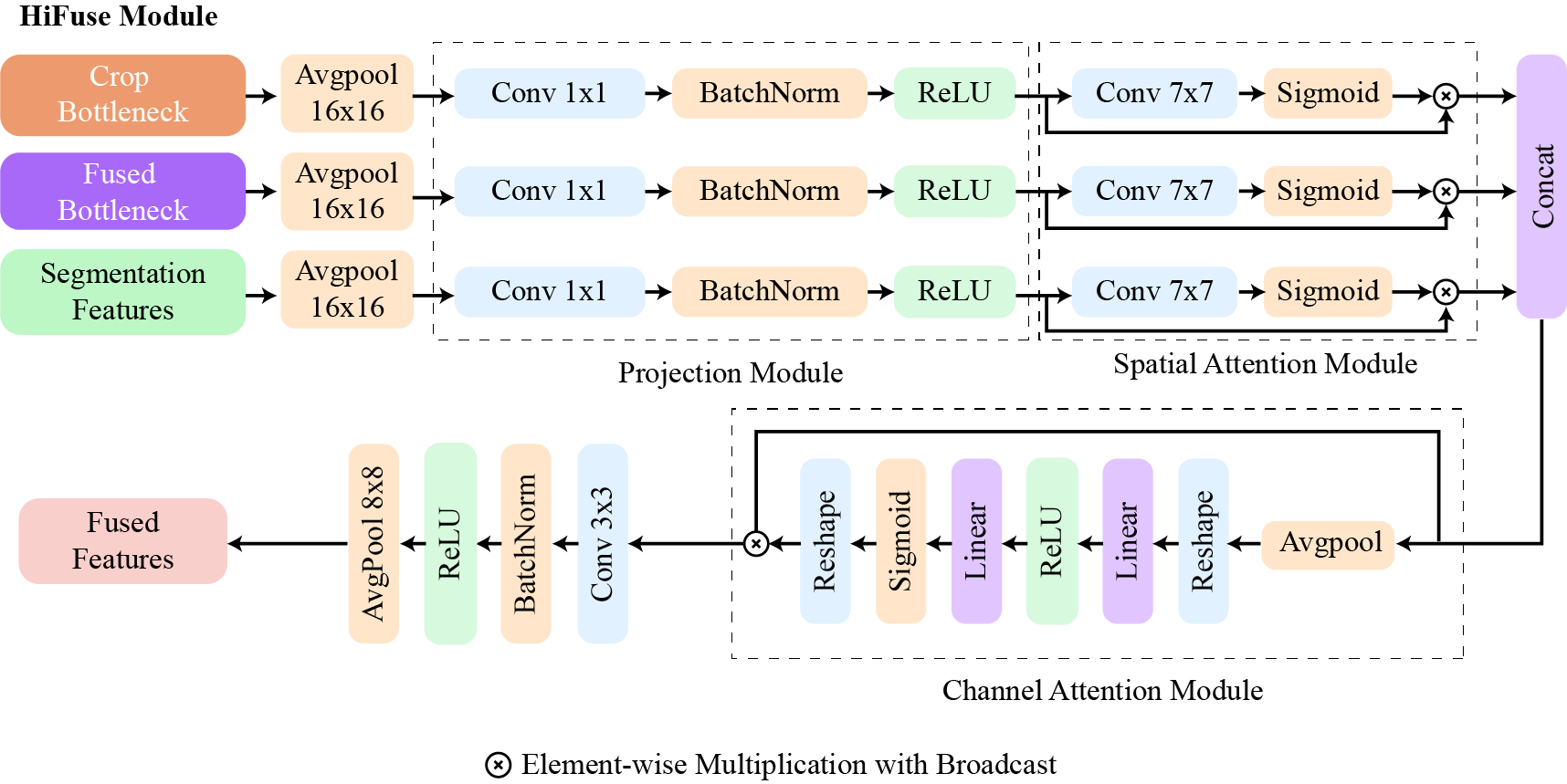}
\caption{Hierarchical multi-scale feature fusion (HiFuse) module.}
\label{fig:hifuse}
\end{figure*}

While the fused bottleneck representation is used for segmentation through the U-Net decoder, the classification task benefits from aggregating features across multiple scales. Inspired by HiFuse \cite{ref13}, we implement the hierarchical multi-scale feature fusion module shown in Figure~\ref{fig:hifuse}. It combines the following feature representations.

\begin{itemize}
\item
  Crop bottleneck features (\(320\) channels): High-level semantic features from the crop encoder, capturing abstract tumor characteristics.
\item
  Fused bottleneck features (\(512\) channels): The output of cross-modal attention, containing integrated local and global information.
\item
  Final decoder features (\(16\) channels): Low-level features from the last decoder stage, preserving fine-grained texture and boundary information.
\end{itemize}

This three-branch design ensures the classifier receives a comprehensive feature hierarchy spanning from low-level texture to high-level semantics, which has been shown to improve classification accuracy for morphologically complex medical images.

\paragraph{Hierarchical Fusion Strategy}\label{hierarchical-fusion-strategy}

Each of the three feature maps is processed through a dedicated pathway:

\begin{itemize}
\item
  Adaptive Spatial Pooling: All features are resized to a common spatial resolution of 16\ensuremath{\times}16 using adaptive average pooling, regardless of their original size. This standardization enables concatenation while preserving spatial structure.
\item
  Channel Projection: Each feature map is projected to a fixed channel dimension (\(512/3 \approx 170\) channels per branch) using 1\ensuremath{\times}1 convolutions followed by batch normalization and ReLU activation. This balances the contribution of each scale.
\item
  Scale-Specific Spatial Attention: A 7\ensuremath{\times}7 convolutional spatial attention mechanism is applied independently to each projected feature map
\item
  Concatenation and Channel Attention: The three attended feature maps are concatenated along the channel dimension (yielding 512 channels total), then processed through a Squeeze-and-Excitation (SE) module \cite{ref22}, which reweights channel importance across scales.
\item
  Final Fusion: The channel-attended features pass through a 3\ensuremath{\times}3 convolution (512 channels), batch normalization, ReLU, and finally adaptive average pooling to 8\ensuremath{\times}8 spatial resolution. The result is a compact 512-channel descriptor summarizing multi-scale information.
\end{itemize}

This hierarchical fusion strategy is a key innovation that enables the classifier to leverage complementary information: bottleneck features provide high-level semantic discrimination (malignant vs. benign patterns), while decoder features preserve fine-grained morphological details (trabecular patterns, margin sharpness) that may be diagnostically relevant.

\subsubsection{Decoder Architecture with Enhanced Attention Gates}\label{decoder-architecture-with-enhanced-attention-gates}

The segmentation decoder follows a symmetric U-Net design \cite{ref23} with four upsampling stages that progressively recover spatial resolution. At each stage, features are upsampled via transposed convolution (stride 2, kernel 2\ensuremath{\times}2) and concatenated with skip connections from the corresponding crop encoder stage. These skip connections are critical for precise boundary localization, as they reintroduce high-resolution spatial details lost during encoding.

\paragraph{Enhanced Attention Gates}\label{enhanced-attention-gates}

Standard U-Net skip connections indiscriminately concatenate encoder and decoder features, potentially introducing noise or irrelevant background information. To address this, we employ attention gates \cite{ref24} that selectively emphasize informative regions of the skip connection based on the decoder\textquotesingle s semantic context. Our enhanced attention gates extend the original formulation by integrating CBAM for additional feature refinement.

At each decoder stage \(i\), let \(g^{i}\) denote the upsampled decoder features and \(\mathbf{x}^{i}\) the skip connection from the encoder. The attention gate computes:

\begin{equation}
\begin{aligned}\mathbf g_{\mathrm{proj}}&=W_g*\mathbf g^{(i)},&\mathbf x_{\mathrm{proj}}&=W_x*\mathbf x^{(i)}\end{aligned}
\end{equation}

\begin{equation}
\psi=\sigma\!\left(W_\psi*\operatorname{ReLU}(\mathbf g_{\mathrm{proj}}+\mathbf x_{\mathrm{proj}})\right)
\end{equation}

\begin{equation}
\mathbf x_{\mathrm{att}}^{(i)}=\mathbf x^{(i)}\odot\psi
\end{equation}

where \(\mathbf{W}_{g}\), \(\mathbf{W}_{\mathbf{x}}\), and \(\mathbf{W}_{\psi}\) are 1\ensuremath{\times}1 convolutions, and \(\sigma\) is the sigmoid activation. The attention coefficient \(\psi \in [ 0,1]^{H \times W}\) acts as a soft binary gate, allowing diagnostically relevant regions to pass through while suppressing irrelevant background.

After attention gating, the attended skip features are further refined using CBAM:

\begin{equation}
\mathbf x_{\mathrm{enhanced}}^{(i)}=\operatorname{CBAM}(\mathbf x_{\mathrm{att}}^{(i)})
\end{equation}

Finally, the enhanced skip features are concatenated with the upsampled decoder features and processed through an EnhancedConvBlock consisting of two 3\ensuremath{\times}3 convolutions with batch normalization, ReLU activations, residual connections, and CBAM attention.

\paragraph{Decoder Channel Progression}\label{decoder-channel-progression}

The decoder mirrors the encoder\textquotesingle s hierarchical structure with progressively decreasing channel counts as spatial resolution increases:

\begin{itemize}
\item
  Stage 4 (deepest): 512 channels \ensuremath{\rightarrow} 112 channels (upsampled from bottleneck)
\item
  Stage 3: 112 channels \ensuremath{\rightarrow} 40 channels
\item
  Stage 2: 40 channels \ensuremath{\rightarrow} 24 channels
\item
  Stage 1 (shallowest): 24 channels \ensuremath{\rightarrow} 16 channels
\end{itemize}

This channel progression ensures sufficient representational capacity at each scale while maintaining computational efficiency. The final 16-channel feature map combines high-resolution spatial detail with multi-scale semantic context, providing an optimal representation for pixel-wise tumor segmentation.

\subsubsection{Multi-Task Output Heads}\label{multi-task-output-heads}

The architecture branches into three prediction heads after feature extraction: a primary segmentation head, a main classification head, and an auxiliary classification head. The main classification head is the one shown after the HiFuse module in Figure~\ref{fig:architecture} and produces the reported three-class prediction; the auxiliary classification head is a separate, much simpler branch that reads the crop encoder bottleneck directly, bypassing cross-modal fusion and hierarchical integration, and exists only to supply an additional gradient path during training. Its output is not used at inference. This multi-task design leverages the complementary nature of segmentation and classification objectives---segmentation requires precise spatial localization and boundary delineation, while classification demands high-level semantic understanding of tumor type. Joint training has been shown to improve both tasks through shared feature learning and implicit regularization.

\paragraph{Segmentation Head}\label{segmentation-head}

The final decoder features (16 channels, full input resolution) are passed through a 1\ensuremath{\times}1 convolution to produce a single-channel logit map representing per-pixel tumor presence probability. To further refine the segmentation output and suppress spurious activations, we apply an additional spatial attention mechanism:

\begin{equation}
\mathbf M_{\mathrm{seg}}=\operatorname{Conv}_{1\times1}(\mathbf F_{\mathrm{decoder}}^{\mathrm{final}})\in\mathbb R^{B\times1\times H\times W}
\end{equation}

\begin{equation}
\mathbf A_{\mathrm{spatial}}=\sigma(\operatorname{Conv}_{7\times7}(\mathbf M_{\mathrm{seg}}))
\end{equation}

\begin{equation}
\mathbf M_{\mathrm{seg}}^{\mathrm{final}}=\mathbf M_{\mathrm{seg}}\odot\mathbf A_{\mathrm{spatial}}
\end{equation}

This spatial attention operates directly on the segmentation logits, computing a 7\ensuremath{\times}7 convolutional attention map that emphasizes spatially coherent tumor regions while suppressing isolated false positives. The use of a relatively large 7\ensuremath{\times}7 receptive field enables the attention mechanism to capture local spatial context and enforce smooth, anatomically plausible segmentation boundaries.

The final segmentation output is obtained by applying sigmoid activation during inference: \({\widehat{\mathbf{y}}}_{\text{seg}}= \sigma(\mathbf{M}_{\text{seg}}^{\text{final}})\), producing pixel-wise probabilities in \([ 0,1]\). A threshold of 0.5 is used to generate binary masks for evaluation.

\paragraph{Main Classification Head}\label{main-classification-head}

The classification head operates on the hierarchically fused features produced by the HiFuse module (Section 2.3.4). After the HiFuse module outputs 512-channel features at 8\ensuremath{\times}8 spatial resolution, global average pooling aggregates spatial information into a 512-dimensional feature vector:

\begin{equation}
\begin{aligned}\mathbf z_{\mathrm{global}}={}&\operatorname{GAP}\bigl(\operatorname{HiFuse}(\mathbf F_{\mathrm{crop}}^{\mathrm{btl}},\\[-2pt]&\mathbf F_{\mathrm{fused}}^{\mathrm{btl}},\mathbf F_{\mathrm{decoder}}^{\mathrm{final}})\bigr)\in\mathbb R^{B\times512}\end{aligned}
\end{equation}

This global descriptor is then processed through a three-layer multi-layer perceptron (MLP) with ReLU activations and dropout regularization:

\begin{equation}
\mathbf z_1=\operatorname{Dropout}_{0.5}(\operatorname{ReLU}(\operatorname{Linear}_{512\to256}(\mathbf z_{\mathrm{global}})))
\end{equation}

\begin{equation}
\mathbf z_2=\operatorname{Dropout}_{0.5}(\operatorname{ReLU}(\operatorname{Linear}_{256\to128}(\mathbf z_1)))
\end{equation}

\begin{equation}
\widehat{\mathbf y}_{\mathrm{class}}=\operatorname{Linear}_{128\to3}(\mathbf z_2)
\end{equation}

The output \({\widehat{\mathbf{y}}}_{\text{class}}\in\mathbb{R}^{B \times 3}\) represents unnormalized logits for the three tumor classes (osteosarcoma, osteochondroma, other benign). During inference, softmax activation produces class probabilities.

The relatively high dropout rate (0.5) serves two purposes: (1) it prevents overfitting to the limited training data by forcing the network to learn robust, redundant representations, and (2) it acts as an implicit ensemble by averaging predictions across multiple sub-networks during training. This aggressive regularization is particularly important for medical imaging, where models must generalize across diverse patient populations and imaging conditions.

\paragraph{Auxiliary Classification Head}\label{auxiliary-classification-head}

In addition to the main classifier, we introduce an auxiliary classification head that operates directly on the crop encoder\textquotesingle s bottleneck features, bypassing the cross-modal fusion and hierarchical integration. This auxiliary head consists of global average pooling followed by a single linear layer:

\begin{equation}
\mathbf z_{\mathrm{aux}}=\operatorname{GAP}(\mathbf F_{\mathrm{crop}}^{\mathrm{btl}})\in\mathbb R^{B\times320}
\end{equation}

\begin{equation}
\widehat{\mathbf y}_{\mathrm{aux}}=\operatorname{Linear}_{320\to3}(\mathbf z_{\mathrm{aux}})
\end{equation}

The auxiliary classifier serves multiple purposes in the multi-task learning framework:

Additional Gradient Signal: It provides a direct gradient pathway from the classification loss to the early layers of the crop encoder, mitigating vanishing gradients and improving feature learning in the encoder backbone. This is conceptually similar to auxiliary classifiers in GoogLeNet \cite{ref25} and other deeply supervised architectures.

Regularization: The auxiliary loss acts as a regularizer, encouraging the bottleneck features to be discriminative for classification even before fusion with global context. This prevents the encoder from learning features that are solely optimized for segmentation at the expense of classification.

Ensemble Effect: During inference, predictions from both the main and auxiliary classifiers can be ensembled (e.g., by averaging logits or probabilities) to potentially improve classification robustness, though we report only main classifier performance in this study.

The simpler architecture of the auxiliary head (single linear layer vs. three-layer MLP) reflects its supporting role---it provides a training signal without introducing excessive parameters that could lead to overfitting.

\subsubsection{Deep Supervision Strategy}\label{deep-supervision-strategy}

Deep supervision \cite{ref26} has been widely adopted in medical image segmentation to improve gradient flow and feature learning at intermediate network layers. We implement deep supervision by attaching auxiliary segmentation heads to three intermediate decoder stages (stages 4, 3, and 2, progressing from coarsest to finest resolution):

\begin{equation}
\mathbf M_{\mathrm{deep}}^{(i)}=\operatorname{Conv}_{1\times1}(\mathbf F_{\mathrm{decoder}}^{(i)})\in\mathbb R^{B\times1\times H_i\times W_i}
\end{equation}

where \(i \in\{ 2,3,4\}\) indexes the decoder stage, and \(H_{i}\times\mathbf{W}_{i}\) is the spatial resolution at that stage. Each auxiliary output is upsampled to the full input resolution using bilinear interpolation to enable direct comparison with the ground truth mask:

\begin{equation}
\mathbf M_{\mathrm{deep,up}}^{(i)}=\operatorname{Interpolate}(\mathbf M_{\mathrm{deep}}^{(i)},\mathrm{size}=(H,W))
\end{equation}

These auxiliary segmentation outputs are supervised using the same loss function as the main segmentation head (detailed in Section 2.4), but with progressively decreasing weight coefficients:

\begin{itemize}
\item
  Stage 4 (lowest resolution, 16\ensuremath{\times}16): weight = 0.2
\item
  Stage 3 (32\ensuremath{\times}32): weight = 0.2
\item
  Stage 2 (64\ensuremath{\times}64): weight = 0.1
\end{itemize}

The total deep supervision loss is computed as a weighted sum:

\begin{equation}
\mathcal L_{\mathrm{deep}}=\sum_{i\in\{2,3,4\}}w_i\,\mathcal L_{\mathrm{seg}}(\mathbf M_{\mathrm{deep,up}}^{(i)},\mathbf y_{\mathrm{mask}})
\end{equation}

where \(w_{i}\) are the weights above and \(\mathcal{L}_{\text{seg}}\) is the segmentation loss function (Unified Focal Loss, described in Section 2.4.2).

The deep supervision loss is then combined with the main segmentation loss:

\begin{equation}
\mathcal L_{\mathrm{seg}}^{\mathrm{total}}=\mathcal L_{\mathrm{seg}}(\mathbf M_{\mathrm{seg}}^{\mathrm{final}},\mathbf y_{\mathrm{mask}})+0.4\,\mathcal L_{\mathrm{deep}}
\end{equation}

This weighting scheme (0.4 coefficient for deep supervision) ensures that the primary objective remains optimizing the final segmentation output, while the auxiliary signals provide supplementary gradient information to intermediate layers. The weights decrease at higher resolutions (stages 3 and 2) because features at these stages are closer to the final output and already receive strong gradients; conversely, deeper layers (stage 4) benefit more from explicit supervision.

Deep supervision provides several benefits:

\begin{itemize}
\item
  Improved gradient flow: Alleviates vanishing gradients in the decoder by providing direct error signals to intermediate layers
\item
  Multi-scale feature learning: Encourages decoder features at each scale to be discriminative for segmentation, improving the quality of skip connections
\item
  Regularization: Acts as an implicit regularizer by enforcing consistency across multiple spatial scales
\item
  Faster convergence: Accelerates training by providing richer learning signals throughout the network depth
\end{itemize}

\subsubsection{Computational Efficiency and Clinical Deployment Analysis}\label{computational-efficiency-and-clinical-deployment-analysis}

The architecture of OsteoHiFuse-Net is specifically designed to balance high-diagnostic sensitivity with the computational efficiency required for real-time clinical deployment. To justify its feasibility, the framework incorporates several parameter-efficient strategies:

Several design choices keep the dual-stream framework within a computational budget compatible with clinical deployment. DenseNet121 was chosen for both encoding streams because its dense connectivity yields efficient parameter utilisation and feature reuse, giving high representational capacity without the parameter counts of wider or deeper backbones. Within the progressive feature enhancement modules, depthwise separable convolutions factorise the standard 3 \ensuremath{\times} 3 operations and reduce computational cost while retaining expressive power. The attention mechanisms are deliberately lightweight: CBAM adds spatial and channel-wise focus at every stage for under a 1\% increase in parameters, and the auxiliary classification head is a single linear layer that supplies a training signal without adding meaningful latency. The decoder reduces its channel count from 112 to 16 as spatial resolution increases, so that the most expensive high-resolution operations act on the fewest feature maps. At inference, the YOLOv11 detector performs first-stage localisation in real time, so ROI extraction does not become a bottleneck in the diagnostic workflow.

\subsection{Loss Function Design}\label{loss-function-design}

Training the dual-input multi-task architecture requires a carefully designed compound loss function that balances three objectives: primary segmentation, main classification, and auxiliary classification. Each objective addresses specific challenges inherent to bone tumor analysis, including severe class imbalance at both the pixel level (tumor vs. background in segmentation) and the sample level (malignant vs. benign cases in classification), as well as small tumor sizes that are easily overlooked by standard loss functions.

\subsubsection{Multi-Task Loss Formulation}\label{multi-task-loss-formulation}

The total training objective is a weighted combination of three loss components:

\begin{equation}
\mathcal L_{\mathrm{total}}=\alpha\mathcal L_{\mathrm{seg}}^{\mathrm{total}}+\beta\mathcal L_{\mathrm{class}}+\gamma\mathcal L_{\mathrm{aux}}
\end{equation}

where:

\begin{itemize}
\item
  \(\mathcal{L}_{\text{seg}}^{\text{total}}\) is the total segmentation loss (main + deep supervision)
\item
  \(\mathcal{L}_{\text{class}}\) is the main classification loss
\item
  \(\mathcal{L}_{\text{aux}}\) is the auxiliary classification loss
\item
  \(\alpha\), \(\beta\), \(\gamma\) are loss weighting coefficients
\end{itemize}

Based on empirical tuning and the relative importance of each task in the clinical workflow (where accurate classification guides treatment decisions while segmentation delineates surgical margins), we set:

\begin{itemize}
\item
  \(\alpha = 0.4\)(segmentation weight)
\item
  \(\beta = 0.5\)(main classification weight)
\item
  \(\gamma = 0.1\)(auxiliary classification weight)
\end{itemize}

The loss weighting coefficients are set to \ensuremath{\alpha}=0.4, \ensuremath{\beta}=0.5, and \ensuremath{\gamma}=0.1. The decision to weight classification (\ensuremath{\beta}=0.5) more heavily than segmentation (\ensuremath{\alpha}=0.4) is directly linked to clinical risk management; in the diagnostic workflow, the consequences of a false negative malignancy (misclassifying osteosarcoma as benign) are far more severe than minor inaccuracies in boundary delineation. While segmentation provides necessary spatial information for surgical planning, classification serves as the primary decision node for determining biological behavior and patient prognosis.

\subsubsection{Unified Focal Loss for Segmentation}\label{unified-focal-loss-for-segmentation}

Bone tumor segmentation presents two key challenges: (1) severe pixel-level class imbalance, where tumor pixels may constitute only 1-5\% of the image, causing the model to bias toward the dominant background class; and (2) small tumor regions that contribute minimally to standard loss gradients, leading to under-segmentation or complete omission. To address both challenges simultaneously, we employ Unified Focal Loss \cite{ref15}, which combines the region-overlap sensitivity of Dice loss with the hard-example focusing capability of Focal Loss.

\paragraph{Dice Component}\label{dice-component}

Dice loss directly optimizes the Dice Similarity Coefficient (DSC), the primary evaluation metric for medical image segmentation:

\begin{equation}
\mathcal L_{\mathrm{Dice}}=1-\frac{2\sum_{i=1}^{N}p_i g_i+\epsilon}{\sum_{i=1}^{N}p_i+\sum_{i=1}^{N}g_i+\epsilon}
\end{equation}

where \(p_{i}= \sigma(\mathbf{z}_{i})\) is the predicted probability for pixel \(i\)(obtained by applying sigmoid to logit \(\mathbf{z}_{i}\)), \(g_{i}\in\{ 0,1\}\) is the ground truth label, \(N\) is the total number of pixels, and \(\epsilon =10^{- 8}\) is a smoothing factor to prevent division by zero. Dice loss directly penalizes region-level mismatch rather than pixel-level errors, making it naturally robust to class imbalance. However, Dice loss alone can be insensitive to small errors and may converge slowly on hard examples.

\paragraph{Focal Loss Component}\label{focal-loss-component}

Focal Loss \cite{ref28} addresses class imbalance and hard examples by down-weighting the loss contribution from easy examples (high-confidence correct predictions) and focusing learning on hard misclassified samples:

\begin{equation}
\mathcal L_{\mathrm{Focal}}=-\frac1N\sum_{i=1}^{N}(1-p_t^{(i)})^{\gamma}\log(p_t^{(i)})
\end{equation}

where \(p_{t}^{(i)}=p_{i}\) if \(g_{i}= 1\)(foreground pixel), or \(p_{t}^{(i)}= 1 -p_{i}\) if \(g_{i}= 0\)(background pixel), and \(\gamma\) is the focusing parameter. When \(\gamma = 0\), Focal Loss reduces to standard cross-entropy. Higher \(\gamma\) values increase the relative importance of hard examples. We use \(\gamma = 2.0\), a value empirically shown to work well for medical segmentation \cite{ref15}.

The focal modulation factor \((1 -p_{t})^{\gamma}\) ensures that:

\begin{itemize}
\item
  Easy examples (high \(p_{t}\), e.g., clear background far from tumor): \((1 -p_{t})^{\gamma}\approx 0\), contributing minimal loss
\item
  Hard examples (low \(p_{t}\), e.g., ambiguous tumor boundaries): \((1 -p_{t})^{\gamma}\approx 1\), contributing full loss
\end{itemize}

This focusing effect is crucial for bone tumors, where challenging cases---irregular margins, subtle periosteal reactions, mineralization patterns---define diagnostic difficulty.

\paragraph{Unified Combination}\label{unified-combination}

The Unified Focal Loss combines both components:

\begin{equation}
\mathcal L_{\mathrm{UFL}}=\mathcal L_{\mathrm{Dice}}+\alpha_{\mathrm{focal}}\mathcal L_{\mathrm{Focal}}
\end{equation}

where \(\alpha_{\text{focal}}= 1.0\) balances the two terms. This combination leverages the complementary strengths of both losses: Dice provides region-level optimization insensitive to pixel-count imbalance, while Focal Loss ensures attention to hard examples and boundary regions. Yeung et al. \cite{ref15} demonstrated that this unified formulation consistently outperforms either loss used alone across multiple medical segmentation benchmarks, including tasks with severe class imbalance comparable to bone tumor segmentation.

For computational efficiency and numerical stability, we implement Focal Loss using binary\_cross\_entropy\_with\_logits, which combines sigmoid activation and cross-entropy in a single fused operation, avoiding numerical issues that can arise from separately computing sigmoid probabilities.

\subsubsection{Class-Balanced Focal Loss for Classification}\label{class-balanced-focal-loss-for-classification}

Sample-level class imbalance in bone tumor datasets---where benign tumors substantially outnumber malignant cases---poses a critical challenge for classification. Standard cross-entropy loss treats all classes equally, causing the model to bias toward the majority class (benign tumors) and potentially misclassify rare but clinically critical malignant tumors. To address this, we employ Class-Balanced Focal Loss, which extends Focal Loss with per-class weighting based on class frequency. The segmentation task utilizes Unified Focal Loss, setting the focusing parameter \ensuremath{\gamma}=2.0 to emphasize hard examples such as ambiguous tumor boundaries and subtle periosteal reactions. For classification, a higher \ensuremath{\gamma}=3.0 is employed to address the inherent difficulty and high inter-observer variability associated with bone tumor subtyping.

To prevent overfitting given the complexity of the dual-input architecture, we employ a dropout rate of 0.5 in the classification head. This value was selected to force the network to learn redundant, robust representations and to act as an implicit ensemble during training, which is critical for generalizability across multi-institutional radiographic data.

The Class-Balanced Focal Loss is formulated as:

\begin{equation}
\mathcal L_{\mathrm{CBFL}}=-\frac1N\sum_{i=1}^{N}\alpha_{y_i}(1-p_{y_i})^{\gamma}\log(p_{y_i})
\end{equation}

\emph{\hfill\break
}

where:

\begin{itemize}
\item
  \(\mathbf{y}_{i}\in\{ 0,1,2\}\) is the ground truth class for sample \(i\)
\item
  \(p_{\mathbf{y}_{i}}\) is the predicted probability for the true class (obtained by softmax activation)
\item
  \(\alpha_{\mathbf{y}_{i}}\) is the class-specific balancing weight for class \(\mathbf{y}_{i}\)
\item
  \(\gamma\) is the focusing parameter (set to 3.0 for classification, higher than segmentation due to harder classification task)
\end{itemize}

\paragraph{Class Weight Selection}\label{class-weight-selection}

The class weights are set from the post-augmentation training distribution to balance gradient contributions. In the order osteochondroma, osteosarcoma, and other benign, they are

\[\boldsymbol{\alpha}=[0.8,\,3.0,\,1.0].\]

These weights reflect the clinical importance and sample scarcity of each class:

\begin{itemize}
\item
  Osteochondroma (28.4\% of training data): \(\alpha_{0}= 0.8\)--- slightly down-weighted, as it is adequately represented and radiographically the most distinctive of the three classes
\item
  Osteosarcoma (25.0\% of training data, malignant): \(\alpha_{1}= 3.0\)--- heavily up-weighted due to clinical criticality. Misclassifying osteosarcoma has severe consequences (delayed treatment, poor prognosis), justifying aggressive rebalancing.
\item
  Other Benign (46.6\% of training data): \(\alpha_{2}= 1.0\)--- baseline weight as the majority class
\end{itemize}

The high weight for osteosarcoma (\(\alpha_{1}= 3.0\)) ensures that the model prioritizes learning discriminative features for malignant tumors, even though they represent only 25.0\% of training samples. This is consistent with clinical priorities where sensitivity for malignancy is paramount.

\paragraph{Focusing Parameter}\label{focusing-parameter}

We use \(\gamma = 3.0\) for classification, higher than the \(\gamma = 2.0\) used for segmentation. This stronger focusing reflects that bone tumor classification is inherently more challenging than segmentation---even expert radiologists exhibit substantial inter-observer variability \cite{ref7,ref9}---and thus requires more aggressive emphasis on hard examples. The higher \(\gamma\) down-weights easy examples more severely, forcing the model to concentrate learning effort on morphologically ambiguous tumors.

The combination of class balancing (\(\alpha\) weights) and hard-example focusing (\(\gamma\) modulation) has been shown to significantly improve performance on imbalanced classification tasks \cite{ref28,ref29}, with particular benefits for underrepresented critical classes like malignant tumors.

\subsubsection{Training Configuration and Optimization}\label{training-configuration-and-optimization}

The model was trained on a single NVIDIA GPU (RTX 3090) using PyTorch 2.0 with CUDA 11.8. Training proceeded for a maximum of 20 epochs with early stopping (patience = 15 epochs) based on validation loss. Model checkpoints were saved only when validation performance improved.

The principal training hyperparameters are listed in Table~\ref{tab:training-hyperparameters}.

\begin{table}[!t]
\centering
\caption{Training hyperparameters for OsteoHiFuse-Net.}
\label{tab:training-hyperparameters}
\small
\begin{tabular}{@{}ll@{}}
\toprule
Hyperparameter & Value \\
\midrule
Batch Size & 16 \\
Input Crop Resolution & $256\times256$ \\
Full Input Resolution & $1024\times1024$ \\
Maximum Epochs & 20 \\
Early Stopping Patience & 15 \\
Optimizer & AdamW \\
Base Learning Rate & $1\times10^{-4}$ \\
Weight Decay & $1\times10^{-4}$ \\
Adam $\beta_1,\beta_2$ & 0.9, 0.999 \\
Warmup Epochs & 5 \\
LR Schedule & Cosine annealing \\
Minimum Learning Rate & $1\times10^{-6}$ \\
Gradient Clipping & Max norm = 1.0 \\
Dropout (Classifier) & 0.5 \\
\bottomrule
\end{tabular}
\end{table}

A warmup--cosine annealing schedule increased the learning rate linearly during the first five epochs, then decayed it smoothly to the minimum value, ensuring stable convergence.

To address class imbalance (46.6 \% Other Benign, 28.4 \% Osteochondroma, 25.0 \% Osteosarcoma), weighted random sampling was applied so that each mini-batch contained a balanced mix of classes. The classification loss incorporated class-balanced weights \ensuremath{\alpha} = {[}0.8, 3.0, 1.0{]} for {[}Osteochondroma, Osteosarcoma, Other Benign{]}, emphasizing the malignant class.

Additional regularization included dropout (0.5), weight decay (1 \ensuremath{\times} 10\textsuperscript{-}\textsuperscript{4}), batch normalization, and gradient clipping (max norm = 1.0). Mixed-precision training improved efficiency, providing faster convergence and reduced memory consumption.

\subsection{Evaluation Metrics}\label{evaluation-metrics}

Model performance was evaluated using a comprehensive suite of metrics covering both segmentation accuracy and classification performance. These metrics were chosen to align with clinical requirements and enable comparison with prior work on bone tumor diagnosis.

\subsubsection{Segmentation Metrics}\label{segmentation-metrics}

Dice Similarity Coefficient (DSC): The primary metric for segmentation performance, DSC measures the overlap between predicted and ground truth masks:

\begin{equation}
\operatorname{DSC}=\frac{2|P\cap G|}{|P|+|G|}
\end{equation}

where \(P\) is the set of predicted foreground pixels and \(G\) is the set of ground truth foreground pixels. DSC ranges from 0 (no overlap) to 1 (perfect overlap), with values above 0.7 generally considered good for medical segmentation and above 0.9 considered excellent. DSC is the standard metric reported in medical segmentation challenges and publications, enabling direct comparison with prior work.

Intersection over Union (IoU): Also known as Jaccard index, IoU provides a complementary measure of region overlap:

\begin{equation}
\operatorname{IoU}=\frac{|P\cap G|}{|P\cup G|}
\end{equation}

IoU is more sensitive to false positives and false negatives than DSC (particularly for small regions), providing a stricter evaluation. The relationship between DSC and IoU is: \(\text{DSC}=\frac{2 \cdot\text{IoU}}{1 +\text{IoU}}\), meaning DSC is always higher than IoU for the same segmentation.

Both metrics were computed per-image and then averaged across the validation set to produce mean DSC and mean IoU scores. These metrics were calculated at a binarization threshold of 0.5 applied to predicted probabilities.

\subsubsection{Classification Metrics}\label{classification-metrics}

Accuracy: Overall classification accuracy measures the proportion of correctly classified samples:

\begin{equation}
\operatorname{Accuracy}=\frac{\text{Number of correct predictions}}{\text{Total number of samples}}
\end{equation}

While intuitive and widely reported, accuracy alone can be misleading for imbalanced datasets, as a model that predicts only the majority class can achieve high accuracy. Therefore, we supplement accuracy with per-class metrics.

Per-Class Precision, Recall, and F1-Score: For each tumor class \(c\), we computed:

\begin{equation}
\operatorname{Precision}_c=\frac{TP_c}{TP_c+FP_c}
\end{equation}

\begin{equation}
\operatorname{Recall}_c=\frac{TP_c}{TP_c+FN_c}
\end{equation}

\begin{equation}
\operatorname{F1}_c=\frac{2\operatorname{Precision}_c\operatorname{Recall}_c}{\operatorname{Precision}_c+\operatorname{Recall}_c}
\end{equation}

where \(TP_{c}\)(true positives), \(\mathbf{F}P_{c}\)(false positives), and \(\mathbf{F}N_{c}\)(false negatives) are computed for class \(c\) in a one-vs-rest manner. These metrics provide insight into class-specific performance:

\begin{itemize}
\item
  Precision measures what proportion of predicted class \(c\) samples are actually class \(c\)(specificity to that class)
\item
  Recall measures what proportion of true class \(c\) samples are correctly identified (sensitivity for that class)
\item
  F1-score is the harmonic mean, balancing precision and recall
\end{itemize}

\section{Results and Discussion}\label{results-and-discussion}

\subsection{YOLOv11 ROI Detection Results}\label{bone-tumor-roi-detection-performance}

The YOLOv11x-based detection module demonstrated promising localization capabilities on the BTXRD dataset. Mean average precision at an intersection-over-union threshold of 0.5 (mAP@0.5, where the "@" denotes the IoU threshold at which a detection counts as correct) reached 75.5\%, supporting its potential for automatic ROI extraction. The detector maintained a recall (sensitivity) of 72.5\% with a corresponding precision of 76.4\%, indicating a balanced trade-off between the proportion of tumors successfully detected and the correctness of the positive predictions. Additionally, the model achieved an mAP@0.5:0.95 of 41.2\% (averaged over IoU thresholds from 0.50 to 0.95 in steps of 0.05, the COCO protocol), reflecting the complexity of precise bounding-box localization for malignant lesions which are often small and morphologically irregular. Despite the challenges posed by radiodensity variations, these results support the feasibility of using automated ROI generation to reduce reliance on manual cropping in the analysis pipeline.

Table~\ref{tab:yolo-performance} summarizes the detector's validation results.

\begin{table}[!htbp]
\centering
\caption{YOLOv11x detection performance on the validation set.}
\label{tab:yolo-performance}
\small
\begin{tabular*}{\columnwidth}{@{\extracolsep{\fill}}lc@{}}
\toprule
\textbf{Metric} & \textbf{Value} \\
\midrule
Precision & 0.764 \\
Recall (Sensitivity) & 0.725 \\
mAP@0.5 & 0.755 \\
mAP@0.5:0.95 & 0.412 \\
\bottomrule
\end{tabular*}
\end{table}

\subsection{Tumor Segmentation Results}\label{tumor-segmentation-results}

The segmentation module of OsteoHiFuse-Net achieved robust and anatomically consistent delineation of tumor boundaries across all three diagnostic categories. Table~\ref{tab:segmentation-results} summarizes the per-class and overall metrics obtained on the held-out test split of BTXRD, evaluated using Dice Similarity Coefficient (DSC), Intersection over Union (IoU), Sensitivity, and Specificity.

\begin{table*}[!t]
\centering
\caption{Per-class and overall segmentation performance on the held-out test split.}
\label{tab:segmentation-results}
\small
\begin{tabular*}{\linewidth}{@{\extracolsep{\fill}}lcccc@{}}
\toprule
Tumor Type & Dice Score & IoU Score & Sensitivity & Specificity \\
\midrule
Osteochondroma & 0.882 & 0.800 & 0.915 & 0.916 \\
Osteosarcoma & 0.923 & 0.859 & 0.911 & 0.941 \\
Other Benign & 0.947 & 0.901 & 0.942 & 0.950 \\
\shortstack[l]{Overall\\(sample-weighted)} & 0.896 & 0.821 & 0.919 & 0.924 \\
\bottomrule
\end{tabular*}
\end{table*}

The proposed framework achieved an overall Dice score of 0.896 and IoU of 0.821, which reflects strong alignment between predicted and ground truth tumor regions. Among individual classes, the Dice values were 0.923 for osteosarcoma, 0.882 for osteochondroma, and 0.947 for other benign lesions. The relatively higher value for benign tumors primarily stems from their smoother and well-defined cortical margins, whereas osteosarcoma often exhibits irregular, permeative boundaries that make precise delineation inherently more difficult. The model's ability to maintain a Dice above 0.92 for osteosarcoma suggests that it preserved sufficient lesion extent for diagnostic interpretation despite this morphological complexity.

This pattern aligns with the expected visual heterogeneity in bone tumors. Osteochondroma cases, which frequently appear as dense, exophytic outgrowths, introduce ambiguity at the bone--tumor interface where mineralization and trabecular overlap occur. In contrast, osteosarcoma lesions require the network to segment through cortical disruption and variable matrix opacity, conditions in which contextual understanding of surrounding bone becomes critical. The incorporation of cross-modal attention fusion appears to have mitigated this challenge by allowing information from the full radiograph to guide the decoder in distinguishing tumor invasion from structural overlap.

\begin{figure*}[!t]
\centering
\includegraphics[width=0.88\textwidth,height=0.66\textheight,keepaspectratio]{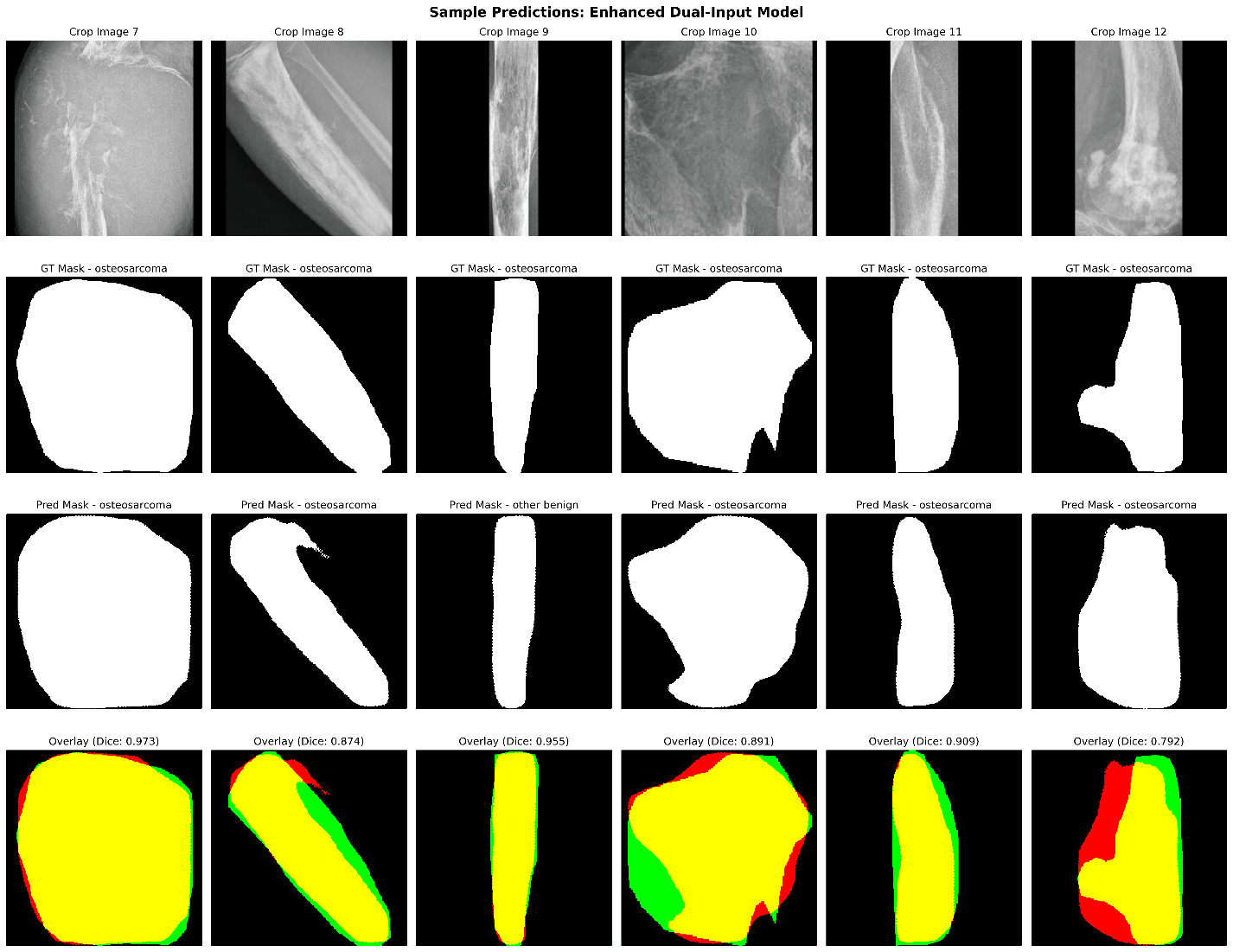}
\caption{Qualitative segmentation performance of the proposed model (green: false negative; red: false positive; yellow: true positive).}
\label{fig:qualitative-segmentation}
\end{figure*}

From a diagnostic perspective, the performance on malignant cases is the most relevant indicator of clinical applicability. The maintained segmentation accuracy for osteosarcoma implies that our model effectively captures features that generalize across diverse radiographic morphologies rather than relying solely on shape regularity. This is important for early detection, as malignant lesions are often present with indistinct or partially ossified boundaries. Figure~\ref{fig:qualitative-segmentation} shows qualitative performance of our model on different classes.

Qualitative examples of the model's predictions show how the segmentation behaves across both different tumor classes. In the overlay maps, green indicates false negatives, red false positives, and yellow true positives. Most of the errors appear around the lesion boundaries rather than within the main tumor area. Overall, the predicted masks remain internally consistent and free from scattered noise, indicating that the network can differentiate real pathological changes from normal bone texture. These observations align with the quantitative results and suggest that the dual-input design mostly produces anatomically coherent and clinically reliable segmentations, even for the irregular shapes typical of malignant tumors.

\subsection{Tumor Classification Results}\label{tumor-classification-results}

The classification branch of OsteoHiFuse-Net was evaluated on the three tumor classes: osteosarcoma, osteochondroma, and other benign lesions. Quantitative results are presented in Table~\ref{tab:classification-results}. The model achieved a macro-averaged F1-score of 0.928 and a macro AUC of 0.993, indicating that it maintained balanced performance across classes despite unequal sample distribution.

\begin{figure*}[!t]
\centering
\includegraphics[width=0.88\textwidth,height=0.66\textheight,keepaspectratio]{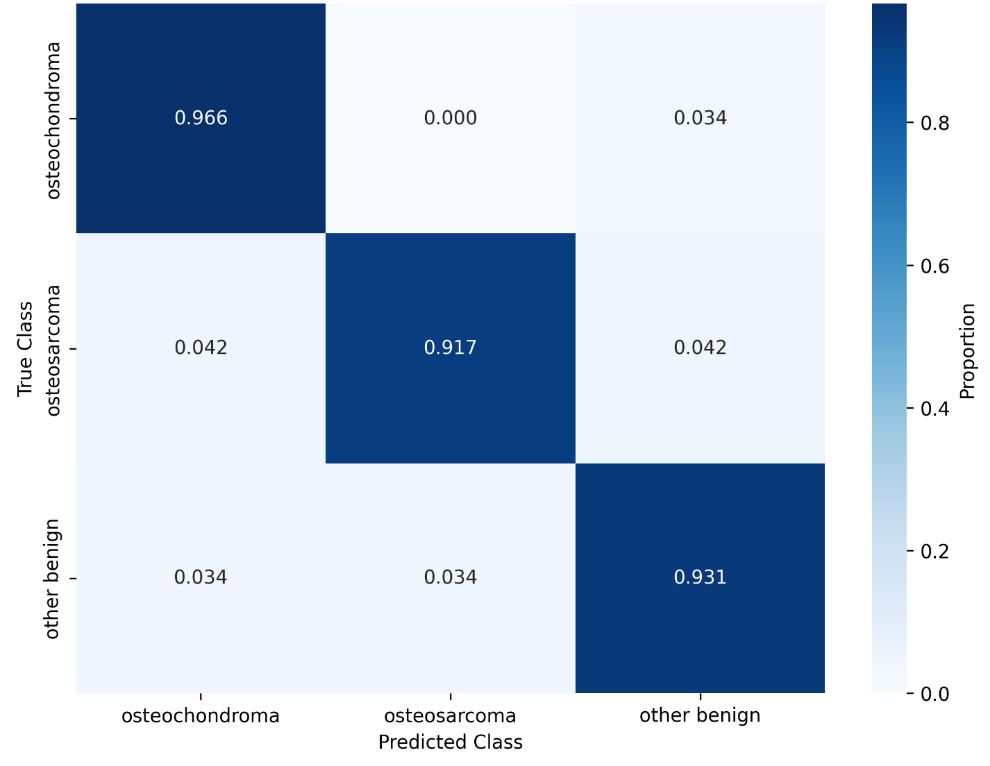}
\caption{Normalized confusion matrix for the three tumor classes.}
\label{fig:confusion-matrix}
\end{figure*}

\begin{table*}[!t]
\centering
\caption{Per-class and averaged classification performance on the held-out test split.}
\label{tab:classification-results}
\footnotesize
\setlength{\tabcolsep}{3pt}
\begin{tabular*}{\linewidth}{@{\extracolsep{\fill}}lcccccc@{}}
\toprule
Tumor Type & Accuracy & Precision & Recall & F1-Score & Specificity & ROC-AUC \\
\midrule
Osteochondroma & 0.965 & 0.986 & 0.966 & 0.976 & 0.962 & 0.993 \\
Osteosarcoma & 0.985 & 0.957 & 0.917 & 0.936 & 0.994 & 0.999 \\
Other Benign & 0.960 & 0.818 & 0.931 & 0.871 & 0.965 & 0.987 \\
Macro average* & 0.955 & 0.920 & 0.938 & 0.928 & 0.973 & 0.993 \\
\bottomrule
\end{tabular*}
\par\smallskip
\begin{minipage}{\linewidth}\footnotesize *The Accuracy entry in this row is the overall (sample-weighted) accuracy; all other entries are macro averages over the three classes.\end{minipage}
\end{table*}

The normalized confusion matrix (\emph{Figure~\ref{fig:confusion-matrix}}) shows that most cases were correctly assigned to their respective categories. Osteosarcoma and osteochondroma were correctly classified in 0.917 and 0.966 of instances, respectively, while other benign lesions reached 0.931. The few misclassifications primarily occurred between osteosarcoma and other benign cases. This overlap can be attributed to the radiographic similarity between certain benign matrix-producing lesions and early-stage malignant forms, where both may display mixed radiodensity and irregular margins.

\begin{figure*}[!t]
\centering
\includegraphics[width=0.88\textwidth,height=0.66\textheight,keepaspectratio]{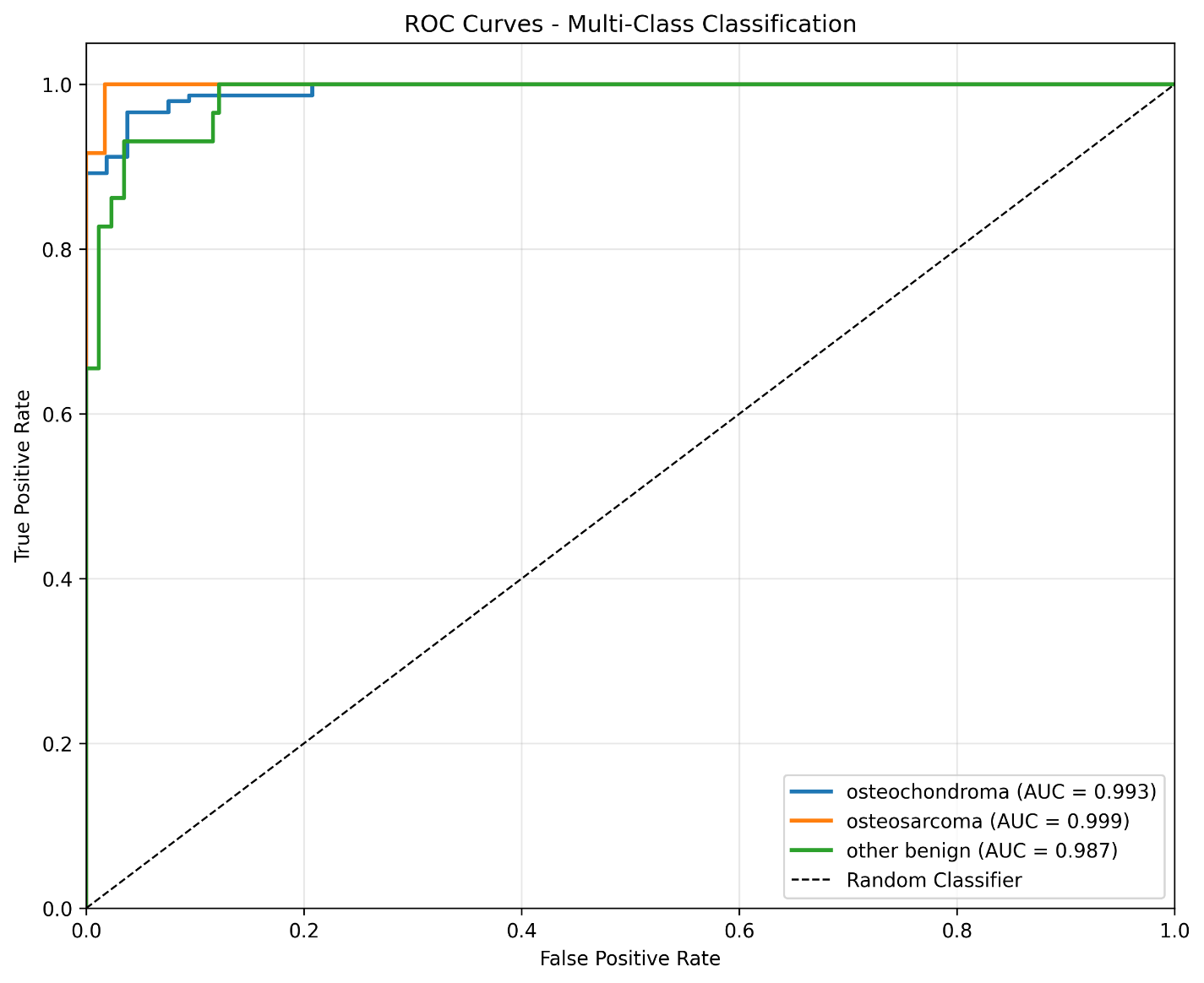}
\caption{Receiver operating characteristic curves for each tumor class (one-vs-rest).}
\label{fig:roc-curves}
\end{figure*}

The curves in \emph{Figure~\ref{fig:roc-curves}} show high separability across all classes, with AUC values of 0.993, 0.999, and 0.987, respectively. The steep initial rise and early plateau near the top-left corner indicate strong true-positive retention with minimal false-positive trade-off across decision thresholds. These findings suggest that the dual-input feature design helped capture both local and contextual patterns relevant to diagnostic differentiation. Malignant cases benefited from the inclusion of full radiograph context, as it provided cues related to periosteal reaction, cortical breach, and medullary expansion. These are features that often extend beyond the cropped region of interest. The high recall and ROC-AUC for osteosarcoma indicate that the model maintained sensitivity to malignancy.

A closer look at class-specific behavior reveals distinct response patterns across diagnostic categories. Osteosarcoma, being the least represented class, showed slightly reduced precision (0.957) compared with benign types but maintained strong recall (0.917), which is preferable in a diagnostic setting where missed malignancies are more consequential than false alarms. The confusion observed toward the benign group generally involved borderline lesions with partial cortical involvement or irregular calcification, which even expert observers can interpret differently depending on projection and image quality. Osteochondroma achieved the highest overall accuracy (0.965) and F1-score (0.976), likely due to its relatively consistent radiographic appearance, characterized by well-demarcated margins and continuity with the host cortex. The classifier appeared to rely on these structural cues rather than local pixel intensity alone, resulting in stable predictions even under varying exposure and orientation conditions. For the other benign category, the model achieved a recall of 0.931 but comparatively lower precision (0.818), indicating occasional misclassification of malignant lesions with smoother boundaries as benign. This suggests that while the network recognized the internal homogeneity typical of benign growths, some aggressive tumors exhibiting low-contrast edges were interpreted as benign-like structures.

In practical terms, these results imply that the framework can reliably distinguish between common benign and malignant bone tumors on radiographs, reducing uncertainty in ambiguous visual presentations. The consistency of predictions across all classes, as reflected in both the numerical metrics and the confusion matrix, indicates that the classifier responded primarily to underlying radiographic structure rather than isolated visual artifacts or dataset bias.

\subsection{Ablation Study}\label{ablation-study}

An ablation analysis was conducted to evaluate the contribution of each major submodule in the proposed framework. Four configurations were compared: (1) a single-branch model using only the full radiograph, (2) the addition of a local crop branch, (3) integration of the CBAM attention modules, and (4) inclusion of the hierarchical fusion (HiFuse) module. Table~\ref{tab:ablation} summarizes the observed changes in Dice score, IoU, classification accuracy, and F1-score.

\begin{table*}[!t]
\centering
\caption{Ablation study over the dual-input, attention and hierarchical-fusion components.}
\label{tab:ablation}
\footnotesize
\setlength{\tabcolsep}{3pt}
\begin{tabular*}{\linewidth}{@{\extracolsep{\fill}}cccccccc@{}}
\toprule
\multicolumn{4}{c}{Submodules} & Dice & IoU & Accuracy & F1-Score \\
\cmidrule(lr){1-4}
Full & Crop & CBAM & HiFuse & Score & Score & & \\
\midrule
$\checkmark$ & & & & 0.862 & 0.781 & 0.884 & 0.730 \\
$\checkmark$ & $\checkmark$ & & & 0.881 & 0.792 & 0.911 & 0.787 \\
$\checkmark$ & $\checkmark$ & $\checkmark$ & & 0.890 & 0.814 & 0.925 & 0.873 \\
$\checkmark$ & $\checkmark$ & $\checkmark$ & $\checkmark$ & 0.896 & 0.821 & 0.955 & 0.928 \\
\bottomrule
\end{tabular*}
\end{table*}

The baseline model trained solely on full radiographs achieved a Dice of 0.862 and IoU of 0.781, indicating adequate segmentation capability but limited precision around irregular tumor borders. Adding the crop branch introduced fine-grained local context, which improved Dice by 1.9 percentage points (0.862 to 0.881) and F1-score by 5.7 percentage points (0.730 to 0.787). This step reduced over-segmentation along trabecular regions by focusing the model's attention on localized morphology.

The integration of CBAM modules further enhanced boundary discrimination and feature recalibration. The network selectively emphasized diagnostically relevant regions, leading to a measurable rise in both segmentation and classification metrics. The observed increase in Dice and accuracy reflects better spatial correspondence and more stable optimization through attention-guided learning.

Finally, the HiFuse module produced the most balanced performance across all metrics. The hierarchical fusion of local and global feature streams enabled consistent segmentation accuracy and improved class differentiation. The full model achieved an overall F1-score of 0.928 and maintained high Dice and IoU scores, indicating coherent lesion boundaries and reliable class-level predictions.

Collectively, these results demonstrate that each submodule contributes progressively to refining the spatial and contextual understanding of the model. The combination of attention-based refinement and hierarchical fusion provides the most stable and generalizable configuration for radiographic bone tumor analysis.

\subsection{Comparison with Existing Literature}\label{comparison-with-existing-literature}

To contextualize the diagnostic performance of OsteoHiFuse-Net, its results were compared with recent deep learning studies on bone tumor classification that reported malignant versus benign discrimination. Table~\ref{tab:literature-comparison} summarizes dataset sizes, classification structures, malignant performance, and imaging modalities.

\begin{table*}[!t]
\centering
\caption{Performance comparison with existing literature.}
\label{tab:literature-comparison}
\small
\textit{(a) Study characteristics}\par\smallskip
\begin{tabularx}{\linewidth}{@{}p{3.1cm}cp{3.0cm}X@{}}
\toprule
Reference & Dataset & Data modality & Classes \\
\midrule
He et al. \cite{ref7} & 1356 & X-Ray & 3 class (benign / intermediate / malignant) \\
von Schacky et al. \cite{ref8} & 934 & X-Ray & Binary \\
Liu et al. \cite{ref9} & 643 & X-Ray + Clinical & 3 class (benign / intermediate / malignant) \\
Song et al. \cite{ref5} & 1305 & X-Ray + CT + Clinical & 3 class (benign / intermediate / malignant) \\
OsteoHiFuse-Net (Ours) & 1613 & X-Ray & 3 class (osteochondroma / osteosarcoma / other benign) \\
\bottomrule
\end{tabularx}
\medskip
\textit{(b) Reported performance}\par\smallskip
\begin{tabularx}{\linewidth}{@{}X >{\centering\arraybackslash}p{3.0cm} >{\centering\arraybackslash}p{3.3cm}@{}}
\toprule
Reference & \shortstack{Multiclass\\accuracy/AUC} & \shortstack{Binary (malignant)\\accuracy/AUC} \\
\midrule
He et al. \cite{ref7} & 0.734/-- & 0.734/0.916 \\
von Schacky et al. \cite{ref8} & -- & 0.802 \\
Liu et al. \cite{ref9} & 0.782/0.872 & 0.782/0.894 \\
Song et al. \cite{ref5} & 0.725/0.847 & 0.725/0.815 \\
OsteoHiFuse-Net (Ours) & 0.955/0.993 & 0.955/0.999 \\
\bottomrule
\end{tabularx}
\end{table*}

The early study by He et al. (2020) trained an EfficientNet-based model on a large multicenter cohort and achieved an overall accuracy of 0.73 for three-way classification. When reduced to a binary malignant-versus-rest task, the AUC increased to 0.92, showing that discriminating malignancy is easier than resolving intermediate lesions. von Schacky et al. (2021) extended this concept through a multitask network combining segmentation and binary classification, achieving 0.80 accuracy on an external test set but limited malignant sensitivity, which highlighted the difficulty of identifying subtle cortical disruptions using single-branch architectures.

Liu et al. (2022) advanced performance by fusing radiographic features with clinical variables such as age and lesion location, reporting a multiclass accuracy of 0.78 and a malignant AUC of 0.89. The gain from multimodal fusion demonstrated the benefit of contextual metadata but required structured data that are not always available in retrospective cohorts. Song et al. (2024) integrated multimodal imaging (X-ray, CT, and MRI) and clinical parameters. Their internal performance reached 0.73 accuracy with an AUC of 0.85, but external testing dropped to 0.63 accuracy and 0.78 AUC, reflecting the challenge of harmonizing multimodal inputs and scanner variability.

In comparison, OsteoHiFuse-Net achieved a multiclass accuracy of 0.955 and macro AUC of 0.993, and in the malignant-versus-rest setting reached 0.955 accuracy and 0.999 AUC using radiographs only. The results suggest that integrating global skeletal context with localized lesion detail can reproduce the discriminative capacity often attributed to multimodal frameworks. The network's attention-guided fusion enables it to capture cortical breach, periosteal reaction, and trabecular irregularity within a single modality, eliminating dependence on auxiliary clinical or tomographic data.

Overall, the comparison indicates that recent progress in bone tumor analysis has shifted from simple CNN-based classification toward models that incorporate spatial context or multimodal information. The performance of OsteoHiFuse-Net demonstrates that when contextual learning is explicitly embedded within radiograph-based architectures, diagnostic precision for malignancy detection can match or exceed that of more complex multimodal systems, offering a more practical approach for deployment in radiographic screening and diagnostic triage.

\section{Limitations and Future Work}\label{limitations-and-future-work}

Although OsteoHiFuse-Net demonstrates strong performance across segmentation and classification tasks, several limitations should be considered when interpreting the results.

First, the detection stage, while outperforming the baseline model reported in the original dataset paper, still leaves room for improvement. The detection module occasionally missed small tumors, particularly those with low contrast or subtle cortical disruption. Since the primary objective of this work was to improve segmentation and classification performance, detection was treated as a preprocessing step rather than an integrated learning objective. Future research could extend the current framework by incorporating detection as part of a unified multi-task architecture, allowing shared feature learning between detection, segmentation, and classification heads. This approach could improve overall consistency and reduce cascading errors from earlier stages.

Second, the dataset contained a wide variety of tumor types distributed across different anatomical locations. However, there were limited examples of each tumor subtype in each location. This imbalance makes it challenging for the model to fully capture the location-specific variation in radiographic appearance, such as differences in cortical structure and background bone density between long bones and flat bones. The limited per-class, per-location representation may therefore affect generalization to unseen tumor types or rare anatomical regions. Expanding the dataset through multi-institutional collaboration and stratified sampling would help alleviate this limitation and enable location-aware model evaluation.

Third, while the dual-input architecture successfully integrates local and global features, it remains relatively complex compared with single-branch networks. The need to process two parallel feature streams increases memory and computational cost, which may limit real-time deployment in low-resource settings. Future optimization efforts could focus on model compression, attention pruning, or transformer-based lightweight fusion strategies to reduce inference time without compromising diagnostic accuracy.

Finally, this study was limited to radiograph-based analysis. Although radiography provides a cost-effective and widely available modality for initial tumor evaluation, it cannot capture certain soft-tissue or intramedullary features that are visible in MRI or CT. A future extension of this work could explore hybrid frameworks that maintain the simplicity of radiograph analysis while selectively incorporating complementary modalities when available.

Overall, future work will aim to integrate detection into the multi-task learning framework, expand and balance the dataset across tumor sites, and explore architectural optimization for clinical deployment. These directions are expected to further enhance the robustness, generalization, and applicability of OsteoHiFuse-Net in real-world diagnostic workflows.

\section{Conclusion}\label{conclusion}

This study introduced OsteoHiFuse-Net, a dual-input framework that integrates local lesion detail with global skeletal context for comprehensive bone tumor analysis on radiographs. The model achieved precise segmentation of both benign and malignant lesions, producing anatomically consistent boundaries even in morphologically complex cases. Quantitatively, it reached a Dice score of 0.896 and IoU of 0.821, while maintaining reliable boundary continuity and minimizing false activations in healthy regions. In classification, the framework achieved an overall accuracy of 0.955 and a malignant AUC of 0.999, demonstrating that well-structured contextual integration within radiographs alone can rival or surpass multimodal systems.

While detection performance and dataset balance remain limitations, the results show that combining spatial context and lesion-specific information within a unified architecture can substantially improve both segmentation and diagnostic interpretation. Future work will focus on integrating detection into the multi-task learning pipeline, expanding dataset diversity across tumor sites, and optimizing model efficiency for real-time clinical use.

\section*{Data Availability}

The Bone Tumor X-ray Radiograph Dataset (BTXRD) analysed in this study is publicly available and was introduced by Yao et al. \cite{ref10}; no new patient data were collected for this work. The class subset, consolidation rules and patient-level split used here are described in full in Sections 2.1 and 2.2.

\section*{Declaration of Competing Interest}

The authors declare that they have no known competing financial interests or personal relationships that could have appeared to influence the work reported in this paper.

\renewcommand{\bibfont}{\small}
\bibliographystyle{elsarticle-num}
\bibliography{ref}

\end{document}